\documentclass{emulateapj}
\usepackage{graphicx}

\def\kms{km~s$^{-1}$} 
\def\dlambda{$\lambda\lambda$}

\slugcomment {Accepted to the Astrophysical Journal}

\shorttitle{High-Velocity Ejecta Jets in Cassiopeia A}
\shortauthors{Fesen \& Milisavljevic}

\begin{document}

\title{An {\sl HST} Survey of the Highest-Velocity Ejecta in Cassiopeia A{\altaffilmark{1}} }

\author{Robert A.\ Fesen}
\affil{6127 Wilder Lab, Department of Physics \& Astronomy, Dartmouth
  College, Hanover, NH 03755, USA}
\and
\author{Dan Milisavljevic}
\affil{Harvard-Smithsonian Center for Astrophysics, Cambridge, MA 02138, USA }

\altaffiltext{1} {Based on observations with the NASA/ESA Hubble Space Telescope, 
obtained at the Space Telescope Science Institute, 
 which is operated by the Association of Universities for Research in Astronomy, Inc.}

\begin{abstract}

We present {\sl Hubble Space Telescope} WFC3/IR images of the Cassiopeia A
supernova remnant that survey its high-velocity, S-rich debris in the NE jet
and SW counterjet regions through [\ion{S}{3}] \dlambda 9069, 9531 and
[\ion{S}{2}] \dlambda 10,287 -- 10,370 line emissions.  We identify nearly 3400
sulfur emitting knots concentrated in $\sim$120${\degr}$ wide opposing streams,
almost triple the number previously known. The vast majority of these ejecta
knots lie at projected distances well out ahead of the remnant's forward blast
wave and main shell ejecta, extending to angular distance of 320$''$ to the NE
and 260$''$ to the SW from the center of expansion. Such angular distances
imply undecelerated ejecta knot transverse velocities of 15,600 and 12,700 km
s$^{-1}$ respectively, assuming an explosion date $\approx$1670 AD and a
distance of 3.4 kpc.  Optical spectra of knots near the outermost tip of the NE
ejecta stream show strong emission lines of S, Ca, and Ar.  We estimate a total
mass $\sim0.1$ M$_{\odot}$ and a kinetic energy of at least $\sim1 \times 10^{50}$
erg for S-rich ejecta in the NE jet and SW counterjet.  Although their
broadness and kinetic energy argue against the Cas~A SN being a jet-induced
explosion, the jets are kinematically and chemically distinct from the rest of
the remnant.  This may reflect an origin in a jet-like mechanism that
accelerated interior material from a Si,S,Ar,Ca-rich region near the
progenitor's core up through the mantle and H,He,N and O-rich outer layers with
velocities that greatly exceeded that of the rapidly expanding photosphere. 

\end{abstract}


\section{Background}

In the early evolution of a supernova remnant, dense clumps of the
highest-velocity ejecta may experience relatively little deceleration and
consequently can catch up and move out ahead of the ever decelerating forward
moving shock wave \citep{Hamilton85}. This situation has been well demonstrated
in laboratory explosion experiments using schlieren and shadow graph techniques
\citep{Settles2006,Biss2007}.  Such clumps interact directly with the
progenitor's circumstellar and interstellar medium and it is this interaction
that allows for their detection.  However, instances of fast outer ejecta in
young supernova remnants are relatively rare.

Although outlying ejecta clumps or ``bullets'' have been seen 
in the Vela supernova remnant \citep{Aschen95,Kat06}, the 
best studied example of SN ejecta located well out ahead of a remnant's forward
blast wave is the young core-collapse remnant Cassiopeia A (Cas A).  With a
current estimated age of around 345 years (SN $\approx$1670:
\citealt{Thor01,Fes06}), Cas A is believed to be the remains of a Type IIb
supernova event \citep{Krause08,Rest08,Rest11,Besel12} probably from a red
supergiant progenitor with an initial mass in the range of 15--25 M$_{\odot}$
\citep{Chevalier03,Young06}.

Outside Cas A's bright main shell of reverse shock heated debris and the
remnant's $4200 - 5200$ \kms \ forward shock front \citep{Delaney03,Patnaude09}
lie thousands of high-velocity ejecta knots. The brightest few of these
outlying ejecta could be seen in the first deep optical images of the remnant
\citep{BM54} and appeared as a faint ``flare'' of emission line knots
extending nearly two arcminutes outside the remnant's northeast rim
\citep{Min59,Min68}.

A follow-up study described the flare feature as a ``jet'' since the emission
knots there were arranged in thin streams in near radial alignment with the
remnant's expansion center \citep{vdb70}.  Any high-velocity emission knots
outside the remnant's northeast limb have subsequently been described as being
part of this NE emission jet ever since.

Proper motion measurements showed the NE jet knots to be the remnant's highest
velocity material and suggested a slightly later explosion date from that
inferred from main shell ejecta measurements (1671 vs.\ 1653: \citealt{Kvdb76}).
Jet knots were also found to be brightest in [\ion{S}{2}] line emission
with weak or absent [\ion{O}{3}] emission \citep{vdb85} unlike the rest of the
remnant's optical emission knots.

Subsequent optical imaging and spectra identified over a hundred
individual knots in the NE jet region \citep{FG96}. Based on radial distances
from Cas~A's expansion center, these ejecta were estimated to have transverse
velocities between 7000 and 13,000 km s$^{-1}$ assuming a remnant distance of
3.4 kpc \citep{Min59,Reed95,Alarie14}.  Northeast jet knots were also found to
show radial velocities of $-4000$ to $+5000$ km s$^{-1}$ with the majority
centered around zero km s$^{-1}$ indicating a jet orientation within a few
degrees of the plane of the sky \citep{Min68,FG96,MF13}.

Deeper optical imaging has revealed the presence of many additional outlying
ejecta knots around most regions of the remnant at projected distances out
ahead of the remnant's forward blast wave seen in X-rays.  Although
initially found only in a small section along the southwest limb and emitting a
[\ion{N}{2}] dominated emission spectrum unlike the sulfur bright NE jet knots
\citep{Fes87}, later surveys of the outskirts of the remnant uncovered dozens
of additional knots all lying out ahead of the X-ray detected blast wave
emission \citep{Fesen01}.

The discovery of a SW ``counterjet'' by \citet{Fesen01} revealed numerous faint
S-rich ejecta knots located far outside Cas~A's southwest limb and positioned
nearly on the opposite side of the remnant from the NE jet region. Confirmation
of extended emission structure off the remnant's SW limb was subsequently
obtained in both X-rays \citep{Vink04,Hwang04} and in the infrared
\citep{Hines04}. 

The existence of a possible NE jet -- SW counterjet arrangement led some to suspect a
bipolar expansion structure for Cas~A
\citep{Fesen01,Hines04,Vink04,Hwang04,Fes06,Schure08}. However, the energy
associated with the jets appeared to be too low to drive the supernova
\citep{Laming06}. 

Moreover, follow-up optical studies suggested the highest velocity S-rich
ejecta were, in fact, arranged in broad fans and not narrow jet-like streams as
first thought \citep{HF08}. Three-dimensional velocity reconstructions of the
remnant's main shell and outlying optical ejecta have established that the NE
and SW jet regions are on truly opposites sides of the remnant and conically
distributed with opening half-angles of at least $\sim 40\degr$ \citep{MF13}.

The exceptionally high velocities of Cas A's outer ejecta through the local
circumstellar medium lead to their gradual disruption and dissolution.
Processes including ejecta knot mass stripping and fragmentation have actually
been observed in Cas~A \citep{Fesen11}, thereby permitting us to witness the
earliest stages of the enrichment process of the interstellar medium with the
products of stellar nucleosynthesis.

A more complete map of the spatial distribution of Cas~A's outer, metal-rich
ejecta knots can provide a better understanding of the dynamics and overall
asymmetries of this core-collapse SN and knot destruction processes.  Toward
these goals, we present here a deep reconnaissance of Cas~A's outlying S-rich
ejecta which can be used to set limits on the number, peak velocities and
kinetic energy of the remnant's highest-velocity ejecta.

\section{Observations}

Previous studies have shown that high-velocity, outlying ejecta are
best studied in the optical and near-infrared. Optical emission from
the northeast and southwest jets can be traced about $90\arcsec$
farther out than in X-rays or infrared, with only a handful of outer
optical ejecta knots around the remainder of the remnant detected in
even the deepest X-ray and mid-infrared images (see \citealt{Fes06}).

Consequently, we undertook a deep near-infrared image survey of Cas A's outer
ejecta using the Hubble Space Telescope's ({\sl HST}) WFC3/IR camera.  These
images reveal a considerably richer and more extensive debris field around the
outskirts of the remnant than previously realized. We also present optical
spectra of a few selected outer ejecta knots with the highest apparent
transverse velocities.  The observations are described in $\S$2 with the
results and discussion presented in $\S$3.  Locations of the detected S
emission knots are shown and marked in several finding charts included in an
Appendix.


\begin{figure*}[ht]
   \centering
   \includegraphics[scale=.90]{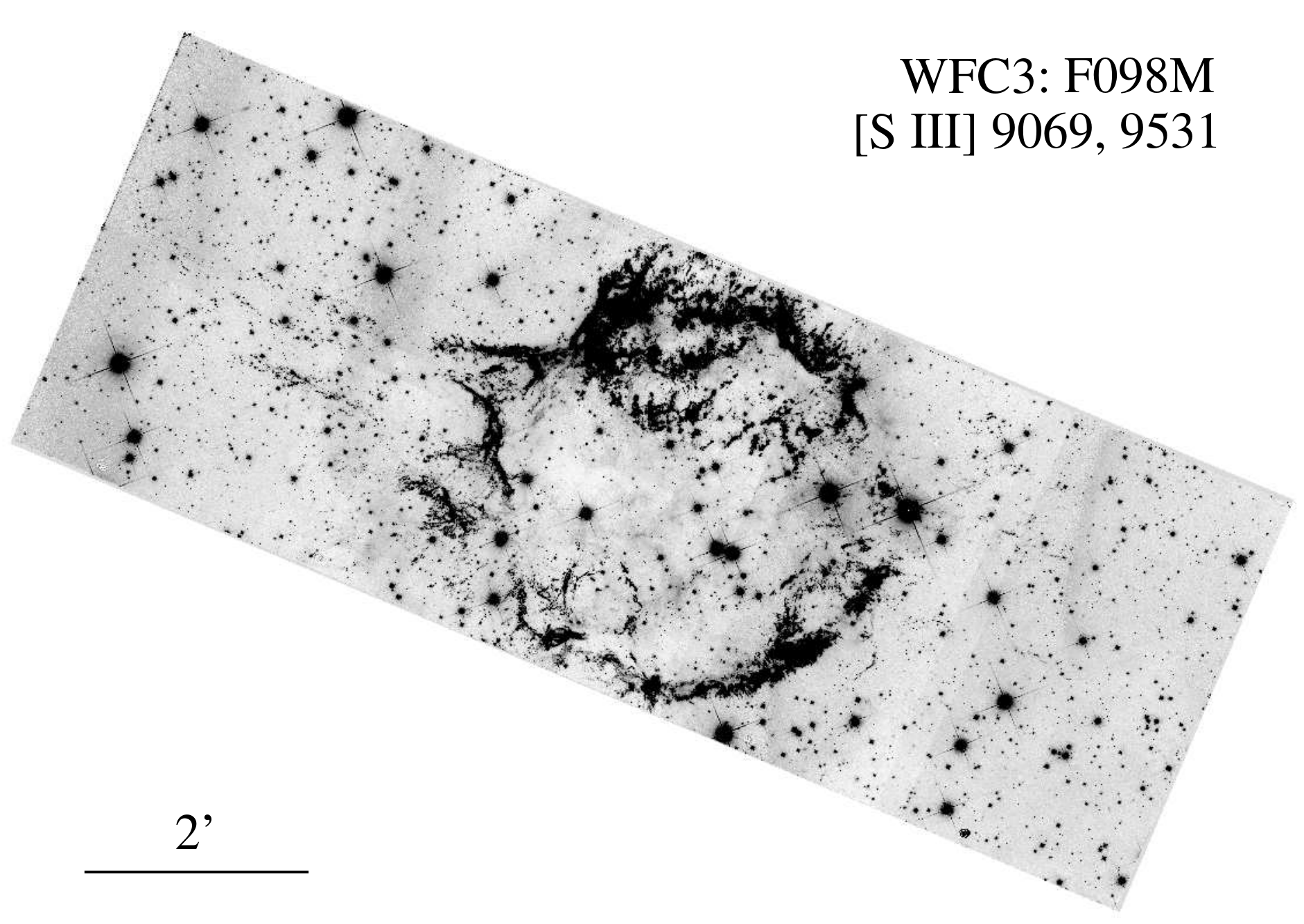}
   \caption{Mosaic of six WFC3/IR F098M images of Cas A obtained in December 2011. }
  \label{fig:introfig}
\end{figure*}


\begin{figure*}[t]
   \centering
   \includegraphics[scale=.90]{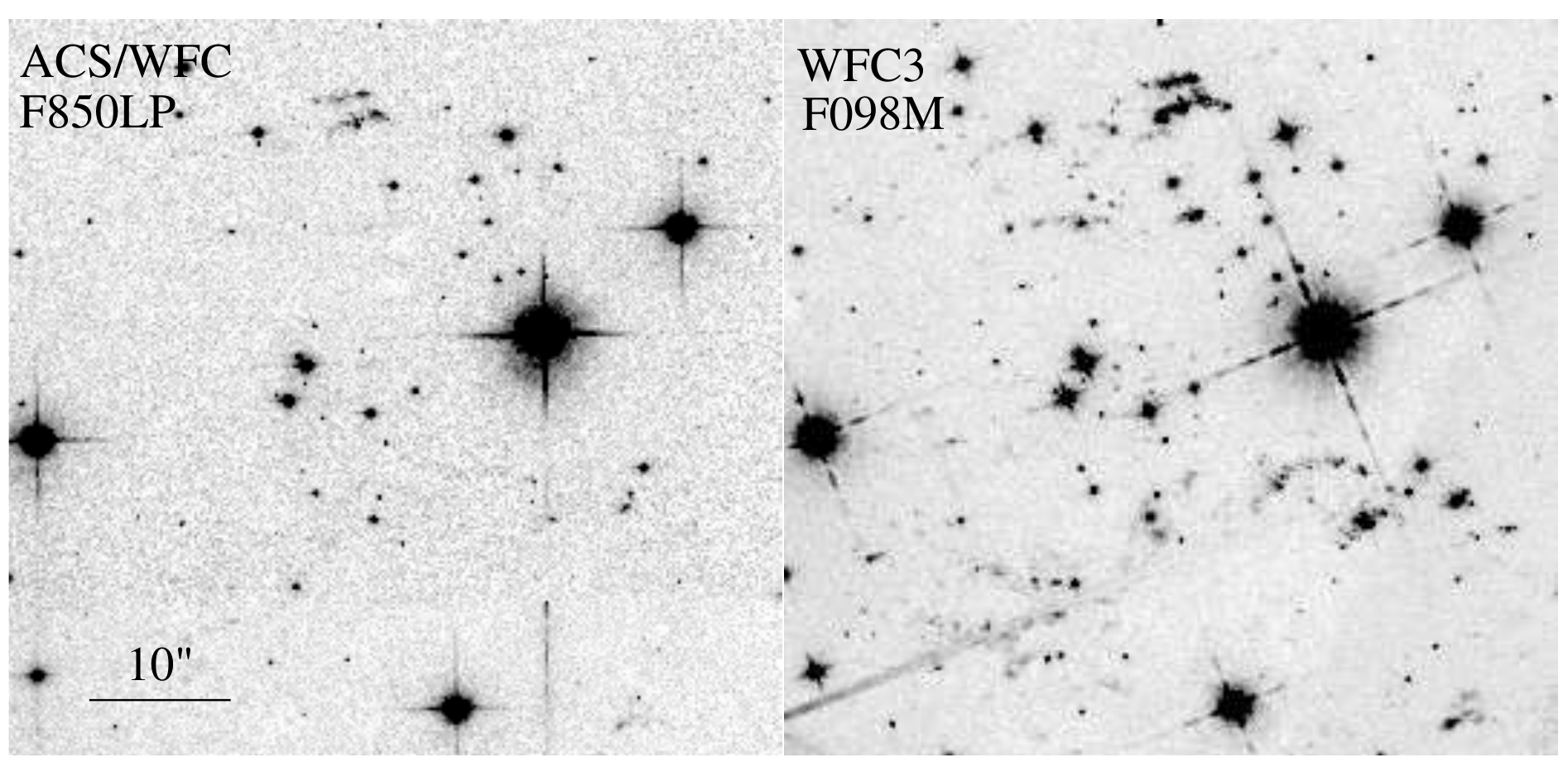}
   \caption{Comparison of ACS F850LP and WFC3/IR images of a small region along Cas A's
            western limb showing the detection of much fainter ejecta
           emission in the WFC3 images compared to ACS images.
           North is up, East to the left. Note the western motion of the ejecta
           features between the December 2004 ACS image and the December 2011 WFC3 image.  }
        \label{fig:ACS_vs_WFC3}
\end{figure*}


\subsection{Near-infrared Image Data}

{\sl HST} images of Cas A were obtained nearly one year apart, on 28 October
2010 and 18 November 2011, using the WFC3/IR camera and the F098M filter. The
IR channel of the WFC3 camera consists of a 1k $\times$ 1k HgCdTe array with a
pixel scale of $0\farcs13$ arcsec pixel$^{-1}$ and a $136 \times 136$ arcsec
field of view.  Six identical, four-point box dithered pointings were used for
both images sets, with a total exposure time for each image set of 22.1 ksec
taken in MULTIACCUM mode during five continuous viewing zone orbits.

The bandpass for the F098M filter + WFC3/IR camera is 9000 to 10700 \AA \ with
an integrated system throughput $\sim$45\%.
The resulting images were primarily sensitive to [\ion{S}{3}]
$\lambda\lambda$9069, 9531 and [\ion{S}{2}] $\lambda\lambda$10,287--10,370
emission lines. We estimate an flux detection limit of these images to be 
$0.55 \times 10^{-18}$ erg cm$^{-2}$ s$^{-1}$.

In order to efficiently cover both NE and SW regions, the images were obtained
with a pattern aligned to the 23\degr\ position angle axis of the NE/SW jets.
This alignment maximized coverage of the remnant's high-velocity NE and SW
streams of outer ejecta knots while also covering the remnant's bright main
shell. 

To assist with the removal of WFC3/IR detector cosmetic defects, the six
image regions were obtained with considerable overlap. Standard pipeline data
reduction of these images was performed using IRAF/STSDAS\footnote{IRAF is
distributed by the National Optical Astronomy Observatories, which is operated
by the Association of Universities for Research in Astronomy, Inc.\ (AURA)
under cooperative agreement with the National Science Foundation. The Space
Telescope Science Data Analysis System (STSDAS) is distributed by the Space
Telescope Science Institute.}.  This included debiasing, flat-fielding,
geometric distortion corrections, photometric calibrations, and cosmic ray and
hot pixel removal.  The STSDAS {\it {drizzle}} task was also used to combine
single exposures in each filter and mosaic them together to form one image.

Figure \ref{fig:introfig} is a mosaic of Cas~A made from the six
WFC3/IR images taken in November 2011 with WFC3/IR + F098M.  Because
the images were taken in Continuous Viewing Zone (CVZ) orbits,
background level in this composite image is not completely uniform due
mainly to Earthshine contamination experienced by some of the six
individual exposures. This resulted in spurious diffuse background
emission in some areas.  Nonetheless, a band of diffuse emission
$\sim$1$'$ off the southeastern limb of the remnant appears to be
real.

\subsection{Optical Spectra}

Ground-based optical spectra of three outlying ejecta knots detected
on these images were obtained in January and December 2012 with the
MMT 6.5 m telescope using the Blue Channel instrument
\citep{Schmidt89}. A 300 l mm$^{-1}$ grating and a $1\farcs25$ slit
were used to obtain spectra of three outer ejecta knots near the
tip of the NE jet.  Spectra have a FWHM of 7~\AA\ resolution
and covered the wavelength region of
$3400-8200$~\AA. Exposures were each 600 sec in duration.

\begin{figure*}[t]
        \centering
        \includegraphics[scale=.75]{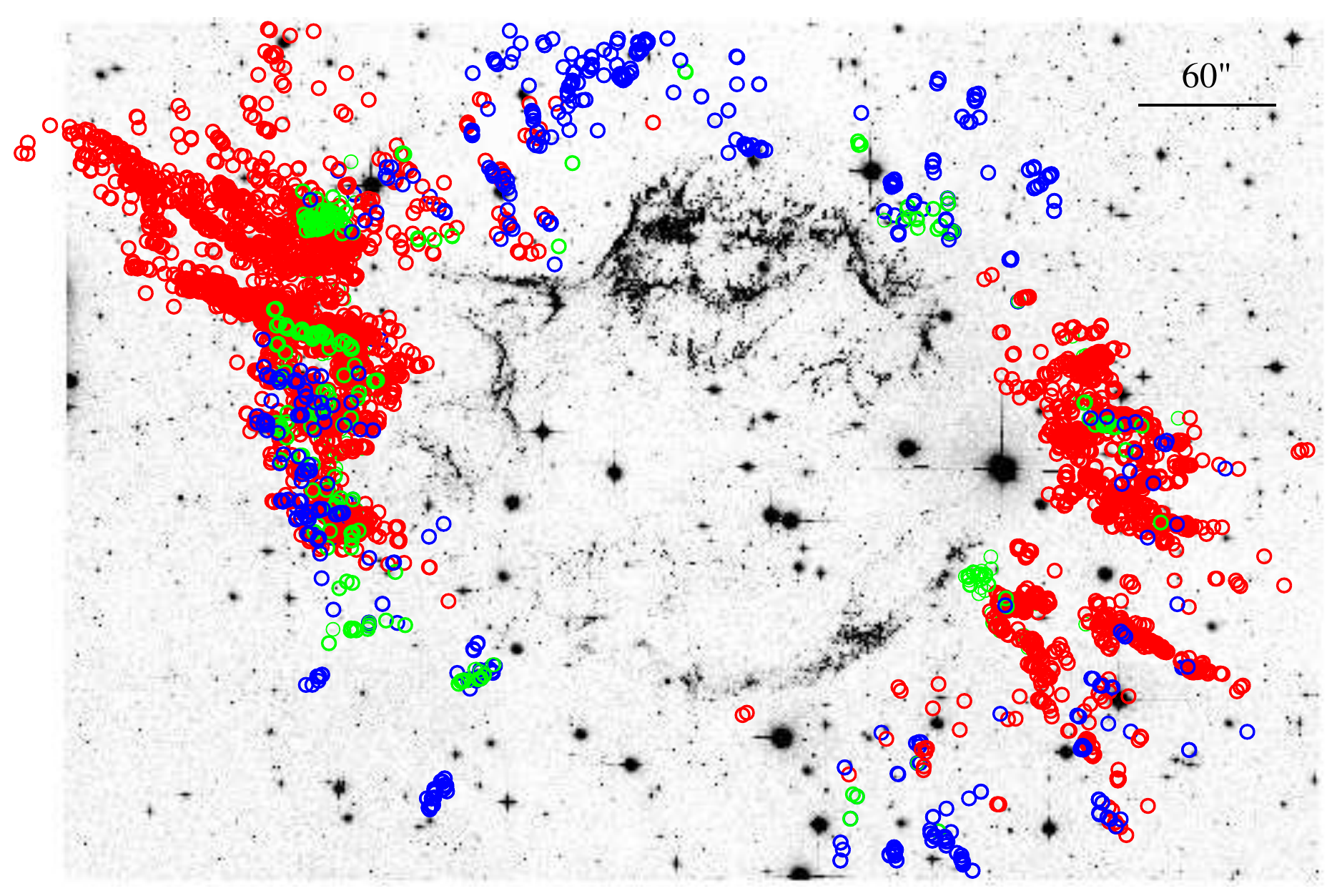}
        \includegraphics[scale=.75]{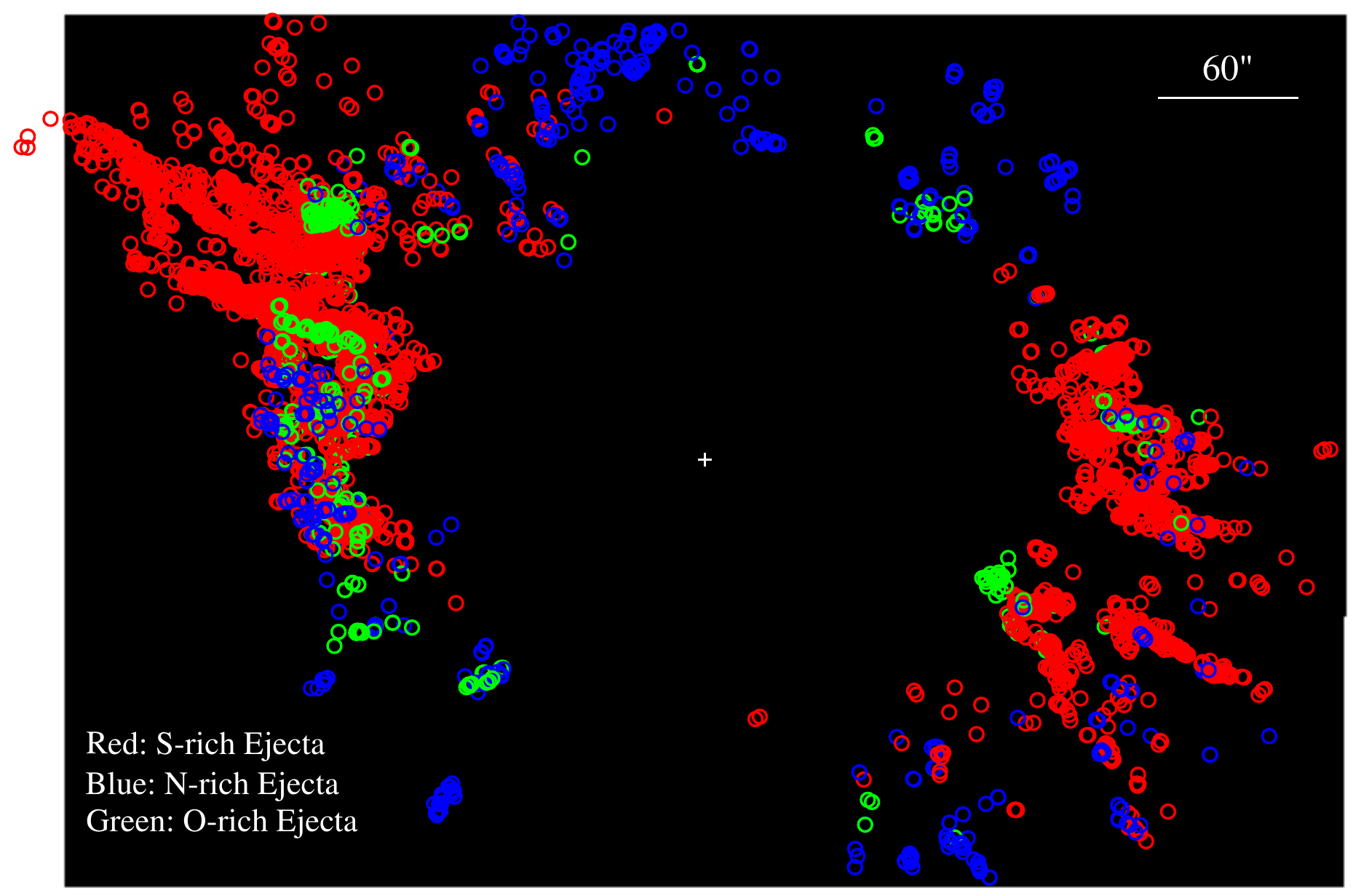}
        \caption{Upper panel: Outer ejecta knot positions shown on the
          December 2004 ACS image.  The blue and green circles
          indicate the 2004.9 locations of [\ion{N}{2}] and
          [\ion{O}{2}] emission knots as defined by \citet{HF08}. The
          red circles mark the December 2011 locations of all knots
          detected in the WFC3/IR camera + F098M filter sensitive to
          [\ion{S}{2}] and [\ion{S}{3}] emissions.  Lower panel:
          Similar ejecta plot as above but now showing the center of
          expansion \citep{Thor01} and the outermost NE ejecta knots
          marking the full extent of the NE ejecta stream.  }
\label{maps_of_ejecta}
\end{figure*}


\begin{figure*}[t]
        \centering
        \includegraphics[scale=.75]{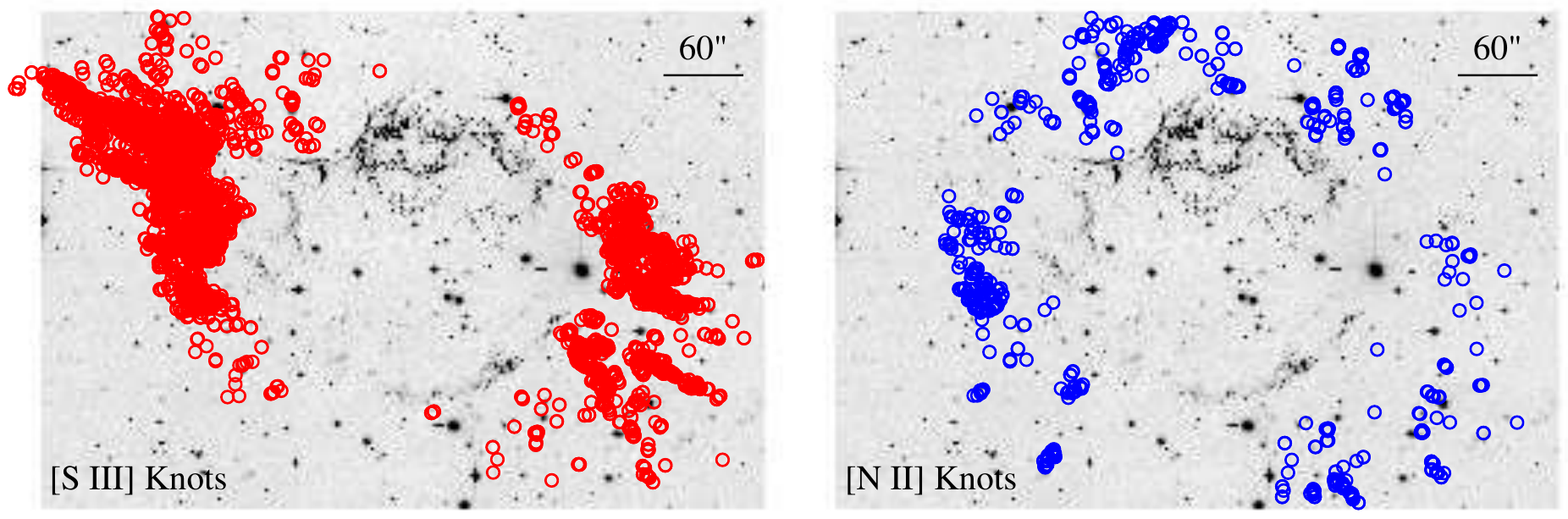}
\caption{Comparison of the projected positions of outer ejecta knots with strong
         [\ion{S}{3}] 9069, 9531 \AA \ emissions (right panel) and
         ones with strong [\ion{N}{2}] 6548, 6583 \AA \ emissions. Note: In order to
         give a more accurate comparison, we have included
         [\ion{S}{2}] 6716, 6731 \AA \ emission knots from
         the \citet{HF08} survey for northern and southern limb regions
         which were not covered by the WFC3 images. }
\label{S_and_N_maps}
\end{figure*}


\begin{figure*}
\centering
\includegraphics[width=0.75\linewidth]{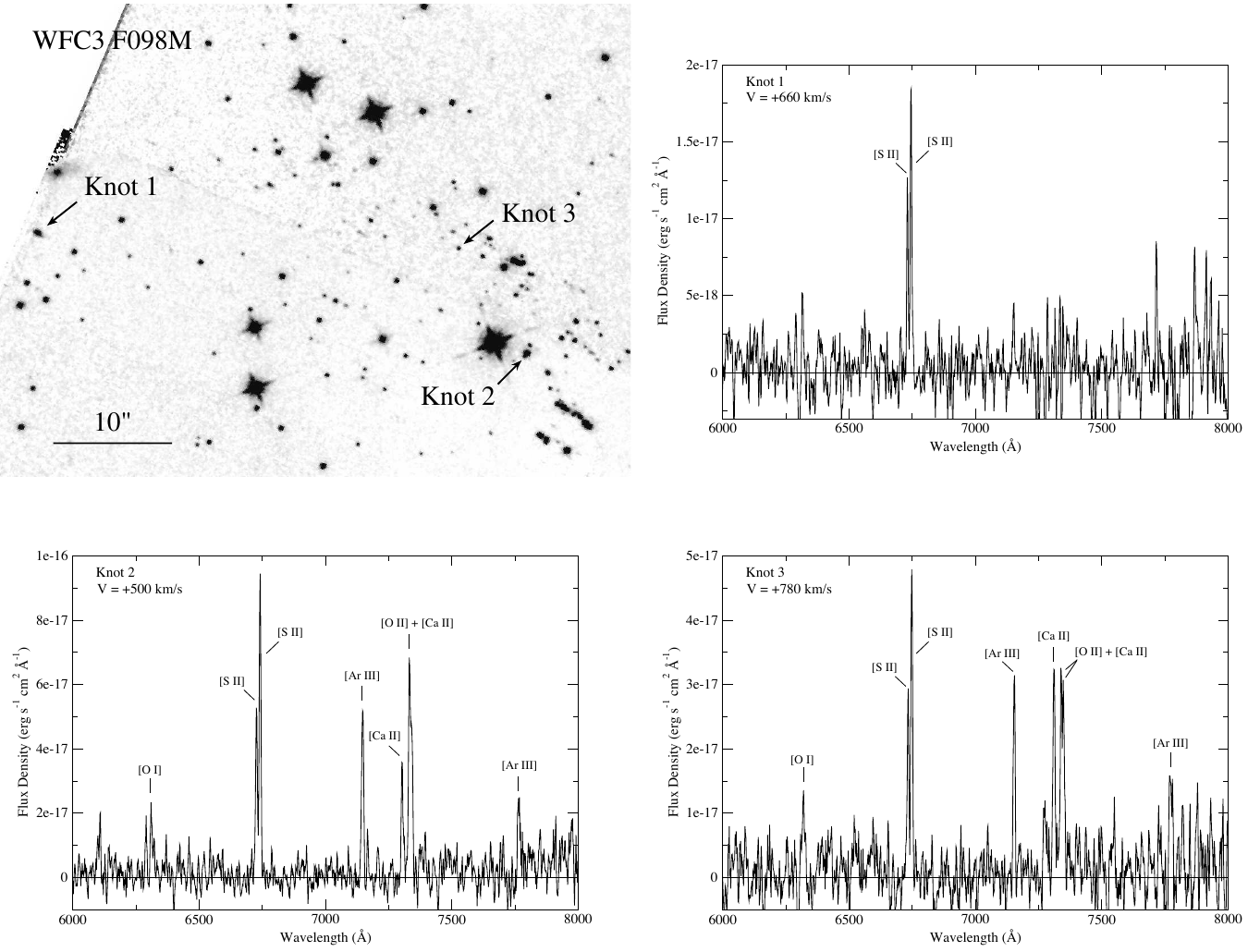}
\caption{Top left: Finding chart showing the location of the three ejecta knots
located at the tip of the NE jet for which we obtained spectra.
Top right and lower panels: spectra of these three NE jet knots. }

\label{fig:knot_spectra}
\end{figure*}

\section{Results and Discussion}

\subsection{The Population of High-Velocity S-rich Ejecta}

The most complete previous survey of Cas A's outer ejecta knots was
that of \citet{HF08} using {\sl HST's} Advanced Camera for Surveys (ACS).
Images were obtained in March and December 2004 using multiple broadband
filters to isolate various emission lines.  They reported finding a
total of 1825 knots.  Of these, filter flux ratios indicated 444 were
dominated by [\ion{N}{2}] 6548, 6583 \AA \ line emissions, 192
dominated by [\ion{O}{2}] 7319,7330 \AA, and 1189 knots with strong
[\ion{S}{2}] 6716, 6731 \AA \ similar to the ``fast-moving knots''
(FMKs) commonly found in the main shell.

Our WFC3/IR F098M images of Cas~A detected much fainter
ejecta knots around Cas A's periphery than the previous ACS F850LP images.
This is shown in Figure \ref{fig:ACS_vs_WFC3} where we compare ACS and
WFC3 images of the same region along Cas A's western limb.  This dramatic
difference is mainly due to the greater sensitivity of WFC3's IR channel to
[\ion{S}{2}] and [\ion{S}{3}] line emission.

Consequently, the WFC3/IR + F098M images uncovered a more extensive debris
field of outer, S-rich ejecta knots than previously realized both in and
outside of the NE/SW jet regions.  We were able to identify 3394 emission
knots, nearly triple the 1189 S-rich FMK-like knots previously cataloged
\citep{HF08}. This number is also roughly an order of magnitude more than 444
and 192 known outer [\ion{N}{2}] and [\ion{O}{2}] knots, respectively
\citep{HF08}. (Note: Finding charts marking the locations of all WFC3 detected
outer ejecta knots are presented in Appendix A.)

Because all outer ejecta knots are only visible through strong interaction with
the local interstellar medium, these $\sim$3400 outer ejecta knots most likely
represent only a fraction of the true population of outer debris fragments. In
addition, the strength of this knot--ISM interaction may vary as the
knot moves through ISM regions of varying density \citep{Fesen11}.

Variability of outer ejecta knots is not uncommon in the NE jet and along the
remnant's eastern limb where high-velocity, outer ejecta are passing through a
highly inhomogeneous ISM or CSM \citep{Fesen11}. Many outlying
ejecta knots have been observed to significantly brighten and fade on
timescales less than one year.

In addition to variable emission flux levels and emission levels below our
detection limit, there is reason to suspect considerably more ejecta knots lie within
the SW jet and along the remnant's western limb than these WFC3 images reveal.
The number of ejecta knots in the NE jet is nearly five times that seen in the
SW jet and this imbalance may have more to do with greater extinction along the
western and southwestern limbs of Cas A due to the presence of a molecular
cloud there \citep{AR96,Keohane96,Kil14} than simply an unequal ejection of material.

\subsection{The Spatial Distribution of Outer S-rich Ejecta}

\citet{MF13} showed that the NE and SW streams of exceptionally high-velocity
ejecta (v $\geq$ 8000 km s$^{-1}$) are arranged in opposing and wide-angle
outflows as initially suggested by their visual appearance in optical images
\citep{Fesen01,Fes06}, in infrared images \citep{Hines04}, and X-ray images
\citep{Vink04,Hwang04,Schure08}.  To explore the nature of these opposing
streams of S-rich ejecta knots, we compared their spatial
distribution to other chemically different outlying high-velocity ejecta.

We begin by addressing differences in the distribution of the S-rich
ejecta versus those that are N-rich or O-rich as defined by relative
emission lines strengths described by \citet{HF08}.  In Figure
\ref{maps_of_ejecta} we show the 2D spatial distributions of S-rich,
N-rich, and O-rich ejecta where the locations of S-rich knots are the
nearly 3400 found from the present WFC3 images while the 444 N-rich
and 192 O-rich knots are those from \citet{HF08}.

As shown in this figure, S-rich knots occupy a much more limited position angle
range while also extending to larger radial distances from the remnant's
expansion center, marked by the small white cross in the lower figure panel.
The farthest S-rich knots extend 320$''$ to the northeast and 260$''$ to the
southwest from the remnant's center of expansion implying undecelerated ejecta
knot transverse velocities of 15,600 and 12,700 km s$^{-1}$ respectively,
assuming an explosion date $\approx$1670 AD and a distance of 3.4 kpc. These
velocities can be compared to maximum transverse velocities based on proper
motions of 13,500 km s$^{-1}$ and 11,500 km s$^{-1}$ for N-rich and O-rich
ejecta, respectively \citep{HF08}.

Particularly striking are the differences in spatial distributions between the three
chemically distinct ejecta types. Figure 4 shows N-rich knots arranged
in a broad shell with two notable gaps in the north and south (for a discussion
of these gaps see \citealt{Fes06}), whereas S-rich knots are concentrated in the NE jet
and SW counterjet regions.  The limited position angle and radial distance of
[\ion{O}{2}] bright knots (Fig.\ 3) has been discussed by \citet{HF08} who
noted increased density of such O-rich knots near the base of the 
NE and SW jet regions.  

\subsubsection{Clumpy Nature of High-Velocity Ejecta}

It is not surprising that the fastest, outermost ejecta in Cas~A are
in the form of small dense clumps.  Ejecta clump formation in
core-collapse supernovae has long been predicted on theoretical
grounds \citep{Gull75,Chevalier78,Kif00,Wang05} and is supported by
considerable observational evidence from both galactic
\citep{Aschen95,Winkler06,MF13} and extragalactic
\citep{Fil89,McCray93,Spy94,Fassia98,Math00,Elm04} supernova and
supernova remnant studies.

Furthermore, several studies have investigated the dynamical interaction when
SN ejecta travel through the surrounding medium
\citep{Hamilton85,Jones94,Anderson94,Cid96,B02,WC02}.  A standard feature of
such ejecta clump/ISM interactions is that a shock is driven into the ejecta
knot, which subsequently undergoes compression and lateral deformation.
Internal shock heating and subsequent cooling via radiative processes
generate substantial X-ray, UV and optical line emissions in ejecta
knots, thus allowing them to be detected.

\subsection{Spectra of the Outermost NE Jet Knots}

Optical spectra of three ejecta knots located in the outer region of the NE
jet are shown in Figure \ref{fig:knot_spectra} along with a finding chart
showing which knots were observed.  As seen in previous spectra of some of the
farthest outlying knots in both the NE and SW jets, two spectral properties of
these outer ejecta are strikingly apparent.

The first property is the remarkably low Doppler velocities observed despite
very large angular distances from the Cas A expansion point. Although the
mid-sections of both jets exhibit radial velocities which can range between
$-4000$ and +5500 km s$^{-1}$ (Fig.\ 7 in \citealt{FG96}), ejecta knots
farthest out have much lower radial velocities, typically under 800 km s$^{-1}$
\citep{FG96,Fesen01,MF13}. 

Our new spectra of the outer knots in the NE jet confirm this finding.
Such low radial velocities for the fastest moving ejecta, whose proper
motions imply transverse velocities well above 12,000 km s$^{-1}$,
indicate an orientation within a few degrees of the plane of the
sky.  The situation is less clear for the SW jet due to far fewer
ejecta spectra \citep{Fesen01,MF13}.  Nonetheless, the few SW knots
with radial distances implying transverse velocities above 12,000
km s$^{-1}$ lie within 10 degrees of the plane of the sky.

The second property of the fastest ejecta knots in the NE jet is the
unusual strength of their [\ion{Ca}{2}] 7291,7324 \AA \ and
[\ion{Ar}{3}] 7136, 7751 \AA \ emission lines relative to [\ion{S}{2}]
6716, 6731 \AA. Out of the hundreds of main shell ejecta observed
\citep{HF96,MF13}, only a small handful of ejecta show comparable
[\ion{Ca}{2}] and [\ion{Ar}{3}] line strengths; the so-called ``strong
Calcium FMKs'' \citep{HF96}.

These S-Ca-Ar outer jet knots also exhibit unusually weak oxygen line emissions
compared to main shell ejecta and have been interpreted as unmixed material
from the thin, Si-Ar-Ca rich layer common in high mass progenitor models.
However, we note that in Knots 2 and 3, some [\ion{O}{2}] 7320, 7330 \AA \ line
emission is also present.

While our spectra are consistent with previously reported outer NE jet knot
spectra \citep{FG96,MF13}, the presence of the [\ion{Ar}{3}] line emissions are
more clearly seen here especially with respect to the weaker [\ion{Ar}{3}] 7751
\AA \ line.  The fact that these S-Ca-Ar knots are the fastest detected ejecta
in the NE jet has been argued as evidence that inner Si, S, Ar, Ca-rich
material was somehow ejected up through overlying material in certain regions,
attaining final outward velocities greater than the progenitor's N and He-rich
surface layers \citep{Fesen01,Fes06,MF13}.

\subsection{Mass and Energy of Outer Ejecta Knots}

Based on the higher resolution ACS images, the average angular size of a bright
S-rich knot is close to $0\farcs1$, with most fainter ones unresolved ($\leq
0\farcs05$). Although we have little direct information about knot emission filling factor, f, but
many of the larger ejecta clumps ($\sim0\farcs4$) show indications of some substructure
below {\sl HST's} image resolution suggesting values less than unity. However, smaller knots
appear nearly stellar suggesting values closer to one.

At a distance of 3.4 kpc, an angular diameter of $0\farcs1$, an
electron density of 4000 cm$^{-3}$ \citep{Fesen01,Fesen11}, an average filling factor, f, of 0.5,  
and a composition of mostly singly and doubly ionized S, Ar, and Ca,
a single S-rich knot would have an estimated mass of $\approx 1 \times 10^{25}$ kg.

We have identified $\sim$3400 S-rich knots from the WFC3 [\ion{S}{2}] +
[\ion{S}{3}] images. Using the above knot mass estimate, this means a total
mass of $\sim0.03$ M$_{\odot}$ for high-velocity S-rich ejecta.  Although much
of the NE jet's emission arises from a few dozen larger knots with angular
diameters of $0\farcs2$ to $0\farcs4$ and hence are the most massive, they are
not numerous enough to significantly change this mass estimate.

However, this mass estimate of outlying ejecta is likely a lower limit since it
pertains only to dense clumps. Extended diffuse material in both jets cannot
be ruled out, plus as noted above there may be hundreds more knots undetected
in our survey in the SW jet along the western limb due to significant
extinction there. This along with possible emission filling factors greater than 0.5
means that the total jet/counterjet mass could be $\sim0.1$ M$_{\odot}$.

Assuming this ejecta mass, one can then estimate the energetics contained in the
remnant's broad NE/SW jet/counterjet flow.  The full range of proper motions
seen in all outer ejecta is $0\farcs45$ to $0\farcs95$ yr$^{-1}$ \citep{HF08}
which translates into transverse velocities of 7000 to 15,000 km s$^{-1}$.
However, the average proper motion is only around $0\farcs6$ corresponding to a
transverse velocity of 10,000 km s$^{-1}$ \citep{HF08}.  Using this average
value, we estimate a total kinetic energy of the NE/SW bipolar flows of 
$1 \times 10^{50}$ (N$_{\rm knots}$/3400) (n$_{\rm e}$/4000 cm$^{-3}$) (f/0.5) erg.
Interestingly, this same energy value was estimated by \citet{Laming06} for
just the NE jet alone based on hydrodynamical models. \citet{Schure08} derived
a lower limit for the jet/counterjet system of $\sim 2 \times 10^{48}$ erg also
from hydrodynamic modeling.

\section{The Nature of the Jet/Counterjet}

The nature and significance of Cas~A's jet/counterjet expansion asymmetry has
long been and remains controversial.  Early on, \citet{Min68} suggested the NE
jet might be the sole surviving part of an outer, high-velocity shell that has
been subsequently decelerated in all other directions. This view was later
supported by \citet{Blondin96} who argued that a jet-like feature of SN ejecta
could be generated in the progenitor's equatorial plane due to pole/equator density
gradients in the local CSM.

However, most researchers now agree that the high-velocity ejecta along the NE--SW limbs
of Cas~A point to a real and significant asymmetrical expansion of the Cas A
supernova \citep{Laming06,Delaney10,Hwang12,MF13}. What they disagree on is the
importance and relevance of the NE/SW jets to the dynamics of the supernova
explosion engine.

Although recently confirmed as being directly opposite of each other and thus
forming a true NE/SW bipolar expansion \citep{MF13}, the possibility of other
high-velocity outflows in the remnant's emission features has raised questions
about the uniqueness of the NE/SW jet/counterjet axis.  A very different
explosion axis has been proposed by \citet{Wheeler08} based on the presence of
Fe-rich ejecta seen outside the remnant's SW limb.  \citet{Delaney10} suggested
that the NE/SW structure is one of several ``pistons'' of high-velocity ejecta,
including an O-Ne rich ejecta alignment along Cas~A's north and south limbs.

\subsection{Evidence for an Explosive Origin}

While there are velocity and abundance variations along Cas~A's main shell,
none match those observed in the NE and SW jet regions. There, ejecta
velocities can reach 15,000 km s$^{-1}$.  In addition, the location of the NE
jet at an obvious rupture point in the remnant's main shell of emission along
with the prominent appearance of the NE and SW jets in X-rays and infrared
images of the remnant has helped to fuel the debate regarding their
significance in understanding the core-collapse mechanism.
 
Several peculiar properties seen only in the NE/SW jet regions also suggest an
explosive origin.  These include: 1) the presence of ``mixed ejecta knots'' that
show a combination of H and N emissions in addition to O, S, and Ar lines, suggesting
a turbulent mixing of photospheric and inner layers, 2) the nearly perpendicular proper
motion of the remnant's central X-ray point source to the NE/SW jet axis \citep{Fes06}, and
3) the unusually weak [\ion{O}{3}] 4959, 5007 \AA \ line emission of jet knots
despite their very high velocities (\citealt{FG96}; Fig.\ 5).

The fact that S-Ar-Ca rich knots are the fastest detected ejecta in the NE jet
has been taken as evidence for locations where inner Si, S, Ar, Ca-rich
material was somehow ejected up through overlying material, attaining final
outward velocities greater than the progenitor's N and He-rich surface layers
\citep{Fesen01,Fes06,MF13}.
Hydrodynamic models by \citet{Laming06} also support the picture where the
NE/SW jets are the result of unusually high-velocity material expelled during
the explosion and not low density regions in the surrounding ISM as proposed by
\citet{Blondin96}. 
Interestingly, \citet{Kil14} recently attributed Doppler
broadened CO emission at millimeter wavelengths from molecular clouds located
nearby and to the southwest of Cas~A to be the caused by interaction between
Cas A's bipolar high-velocity outflows and the molecular gas.

An explosive origin means that our energy estimate of $1 \times 10^{50}$ erg
discussed above may be an underestimate since it does not include the energy
needed to propel jet material up through the progenitor's outer layers.  This
added energy, together with the possibility of additional mass from undetected
diffuse jet material, means the jets likely contain $\sim$10\% or
more of the expected 10$^{51}$ erg supernova explosion energy.

\subsection{The Role of Jets in the Cas A Supernova }

A particularly important aspect of Cas A's jets is their lack of Fe-rich
ejecta.  In jet-induced explosion models, considerable Fe-rich ejecta is
expected to be found in high-velocity jet regions \citep{Khokhlov99}.  However,
no Fe-rich ejecta is seen in either the NE or SW jets, whereas X-ray
observations show considerable Fe-rich ejecta elsewhere in the remnant with
velocities as high as 7000 km s$^{-1}$ \citep{Hughes00,Hwang03}.

Both the NE and SW jets show strong Si enrichment in X-rays \citep{Vink12} and
our optical spectra show that the fastest moving knots near the tip of the NE
jet exhibit unusually strong Ar and Ca emissions. While this may indicate that NE/SW
jet material originates from layers deep inside the progenitor, it is not clear
how such a high-velocity, jet/counterjet bipolar structure fits into the
overall core-collapse explosion dynamics. 

Rather than being signatures of a jet-driven explosion, current observations indicate the
Cas~A supernova explosion was most likely the result of uneven neutrino heating.  Spatial
mapping of radioactive $^{44}$Ti emission appears to rule out a strongly
bipolar explosion and instead favors low-mode convective instability
\citep{Grefenstette14}.  The remnant's bubble-like interior morphology that
smoothly connects with multi-ringed structures seen in the main shell of
expanding debris is consistent with this picture \citep{MF15}. 

Doppler mapping of the X-ray bright Fe-K emission in Cas~A has shown that three
sizable portion of Fe-rich material sits within three of these rings
\citep{Delaney10,MF13}. Thus, Cas~A's large internal cavities of
non-radioactive ejecta may have originated from turbulent mixing processes that
encouraged the development of outwardly expanding plumes of radioactive
$^{56}$Ni-rich ejecta.

\citet{Burrows05} proposed that the morphology and debris pattern of
Cas~A and the NE/SW jets may have an origin in two ``explosions''
taking place at different times: one, the neutrino-driven supernova,
and the other a MHD jet or B-field-modified proto-neutron star wind. In
this picture, the NE/SW outflows would have emerged after the
supernova explosion from the core region and into the already
expanding supernova debris. 

\subsection{A Continuum of Supernovae with Jets}

It is of interest to consider our findings in the context of recent
work on supernovae that have established a continuum of explosion
energies extending from broad-lined Type Ic SNe associated with
gamma-ray bursts, to more ordinary Type Ib/c SNe
\citep{Soderberg10,Chakraborti15}. The existence of such events
suggests that a wide variety of jet activity may potentially be
occurring at energies that are hidden observationally
\citep{Margutti14,Milisavljevic15}. In these cases, the central engine
activity stops and becomes ``choked'' before the jet is able to pierce
through the stellar envelope. Supernovae associated with choked jets
lack sizable amounts of relativistic ejecta and thus can be
dynamically indistinguishable from ordinary core-collapse SNe
\citep{Lazzati12}

Whether or not the jet successfully emerges from the star is dependent
on many factors. Perhaps the most important factor is core rotation at
the time of core collapse. For most supernovae, rotation effects would
be small and the explosion neutrino-driven. However, in a small subset
of cases where core rotation of the presupernova star is rapid,
magnetic fields will be amplified and make magnetohydrodynamic (MHD)
power influential \citep{Akiyama03}. In extreme cases where the
rotation rate is very fast, MHD processes could dominate and a
hypernova and/or a GRB could result
\citep{Woosley06,Burrows07}. Another factor is the progenitor star
size and composition. Large stellar envelopes and/or He layers may
inhibit central jets from completely piercing the surface of the star
(e.g., \citealt{Mazzali08}).

Given that the energy associated with the NE/SW jets of Cas~A is an
order of magnitude below the energy anticipated to be associated with
the original supernova explosion, any observational signature of their
presence would be hidden in SNe at extragalactic distances. Only through
nearby, resolved inspection can we be aware of their existence. In
this context, the jet/counterjet in Cas~A may be interpreted as
evidence for a continuum of jet activity strength in core-collapse
supernova explosions.

\section{Conclusions}

We have presented deep {\sl HST} WFC3/IR images of the Cassiopeia A supernova
remnant in order to survey its highest velocity debris in the NE jet/SW
counterjet regions.  These high resolution, near-infrared images provide the
deepest and most complete inventory of the remnant's highest-velocity,
metal-rich material with unsurpassed detail and depth. 

We conclude the following:

1) Roughly 3400 outlying sulfur emitting knots are detected, which is nearly triple the
previously known number.  The majority of newly detected knots lie at projected
distances well out ahead of the remnant's forward blast wave and main shell ejecta
with estimated transverse velocities approaching $\sim 15,000$ \kms, some 8000
\kms\ more than than the bulk ejecta in the remnant's reverse shock heated main
shell.  Because the jet knots extend out to the edge of the WFC3 detector in
the NE region (Fig.\ 5), the true extent of the outflow is unknown and more material at
even greater velocities may exist.

2) Assuming a significant number of additional ejecta knots lie undetected due
to a lack of strong CSM/ISM interaction leading to detectable shock emission
plus undetected knots in the high extinction SW jet region, we estimate a total
mass of $0.1$ M$_{\odot}$ for the remnant's high-velocity, outlying S-rich
ejecta.  Using this mass and the observed expansion velocities, we estimate
that the kinetic energy imparted to the NE/SW jets was $\sim 1 \times 10^{50}$
erg.  However, this may be an underestimate since it does not include
the energy needed to propel jet material up through the progenitor's outer
layers.  This, together with the possibility of additional mass from undetected
diffuse jet material, means the jets could contain $\sim$10\% or more of the
overall expected 10$^{51}$ erg supernova explosion energy.

3) Optical spectra of ejecta knots near the farthest tip of the NE ejecta
stream are dominated by emission lines of S, Ca, and Ar. These spectra are
significantly different from the main shell ejecta especially in their strength
of [\ion{Ar}{3}] 7136,7751 and [\ion{Ca}{2}] 7291,7324 line emissions. 

4) The NE jet and SW counterjet streams of ejecta are chemically and
kinematically unlike any other observed areas of Cas~A.  They appear to have
formed around the time of core collapse by an explosive, jet-like mechanism
that accelerated interior material from the Si–S–Ar–Ca rich regions near the
progenitor's core up through and out past the He and O rich outer layers with
velocities that greatly exceeded that of the expanding photosphere.

Our findings are generally consistent with the dynamical scenario put forth by
\citet{Burrows05} who proposed that the NE and SW jet/counterjet is associated
with a proto-neutron star wind that emerged after the neutrino-driven supernova
explosion and pushed into the expanding, turbulently mixed supernova debris.
They are also consistent with the notion that a continuum of high-velocity, bipolar
expansions may exist in core-collapse supernovae. 

\acknowledgements We thank D. Patnaude for helpful discussions about Cas A, the
staffs of MDM and MMT Observatories for observing support that made the
ground-based observations possible, C. Black for assistance with the optical
spectra, and the referee for helpful comments. This work was supported by NASA
through grants GO-12300 and GO-12674 from the Space Telescope Science Institute
(STScI), which is operated by the Association of Universities for Research in
Astronomy.  A portion of the observations reported here were obtained at the
MMT Observatory, a joint facility of the Smithsonian Institution and the
University of Arizona.

{}   

\clearpage
\newpage


\section{Appendix A: Outer Ejecta Knot Finding Charts}

The most complete previous survey of high-velocity ejecta outside of Cas A's
main shell of expanding debris is that of \citet{HF08}. They published a
catalog of over 1800 ejecta knots along with finding charts and 2004
coordinates, proper motions, photometric filter fluxes, and estimated knot
emission types based on filter flux ratios. While our survey is based on a
single filter image, specifically the near infrared WFC3/IR F098M filter, these
images revealed significantly more ejecta knots in the outlying regions of Cas
A  in the NE and SE jet regions. 

In this appendix, we present finding charts of the S-rich ejecta knots
visible in one of the WFC3/IR F098M images in the outskirts of Cas A.  For regions
where the ejecta knots are the densest, we include enlargements,
with and without identifying circles, to facilitate
individual knot and ejecta stream identification.

Due to the high proper motions of the outer ejecta ($0.3'' - 0.9''$ yr$^{-1}$)
and the fact that many of these knots exhibit significant brightness changes
over time spans as short as a few months \citep{Fesen11}, we have not included
positional information for these outlying knots. Instead, we view the finding
charts presented here as a more appropriate means of documenting the remnant's
current outer ejecta debris population and distribution.   

Figure 6 shows the ten sections of the  northeast and southwest outer regions
of Cas A where enlargement images are presented in Figures 7 through 11. The
coordinates for the centers of these ten finding charts are listed in Table 1.
Greater image enlargements of particularly dense knot regions are shown in
Figures 12 through 18.

\begin{deluxetable*}{cccccc}
\tablecolumns{6}
\tablecaption{Coordinates for Outer Ejecta Knot Finding Charts} 
\tablehead{
\colhead{Region} &
\colhead{RA (J2000)} &
\colhead{Dec (J2000)} &
\colhead{Region} &
\colhead{RA (J2000)} &
\colhead{Dec (J2000)} \\ 
\colhead{ } & \colhead{ h \ m \ s ~~} & \colhead{~~~  \degr ~~ \arcmin ~~ \arcsec} &
\colhead{ } & \colhead{ h \ m \ s ~~} & \colhead{~~~  \degr ~~ \arcmin ~~ \arcsec} } 
\startdata
 A & 23:23:58.10 & +58:51:17 & F & 23:23:04.28 & +58:49:34 \\
 B & 23:23:43.35 & +58:51:17 & G & 23:23:07.72 & +58:48:11 \\
 C & 23:23:58.05 & +58:49:53 & H & 23:23:59.70 & +58:48:11 \\
 D & 23:23:43.35 & +58:49:53 & I & 23:23:17.09 & +58:46:48 \\
 E & 23:23:47.75 & +58:48:27 & J & 23:23:03.47 & +58:46:48 \\  
\enddata
\label{tab:table1}
\end{deluxetable*}


\begin{figure*}[b!]
        \centering
        \includegraphics[scale=.70]{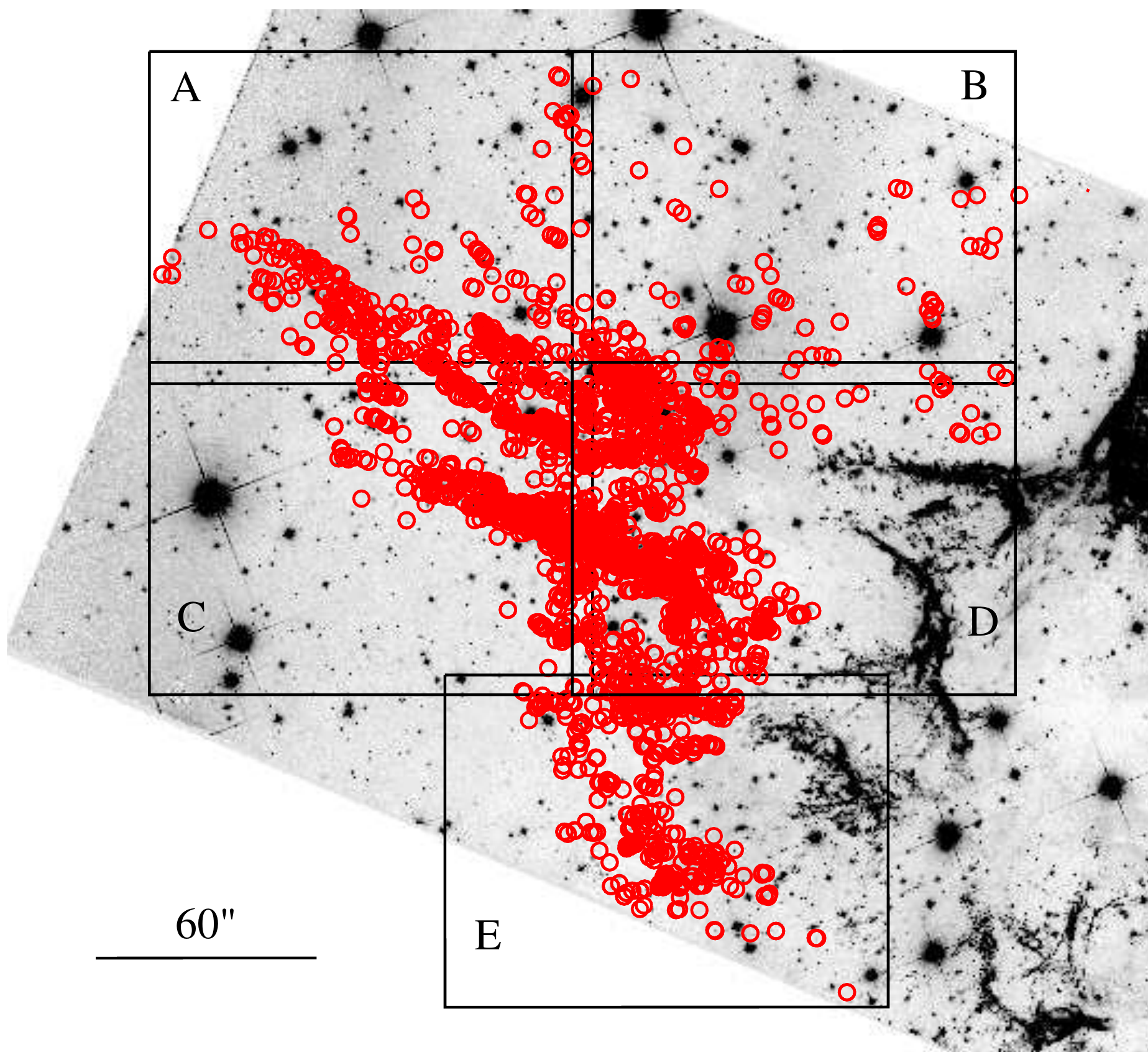}
        \includegraphics[scale=.70]{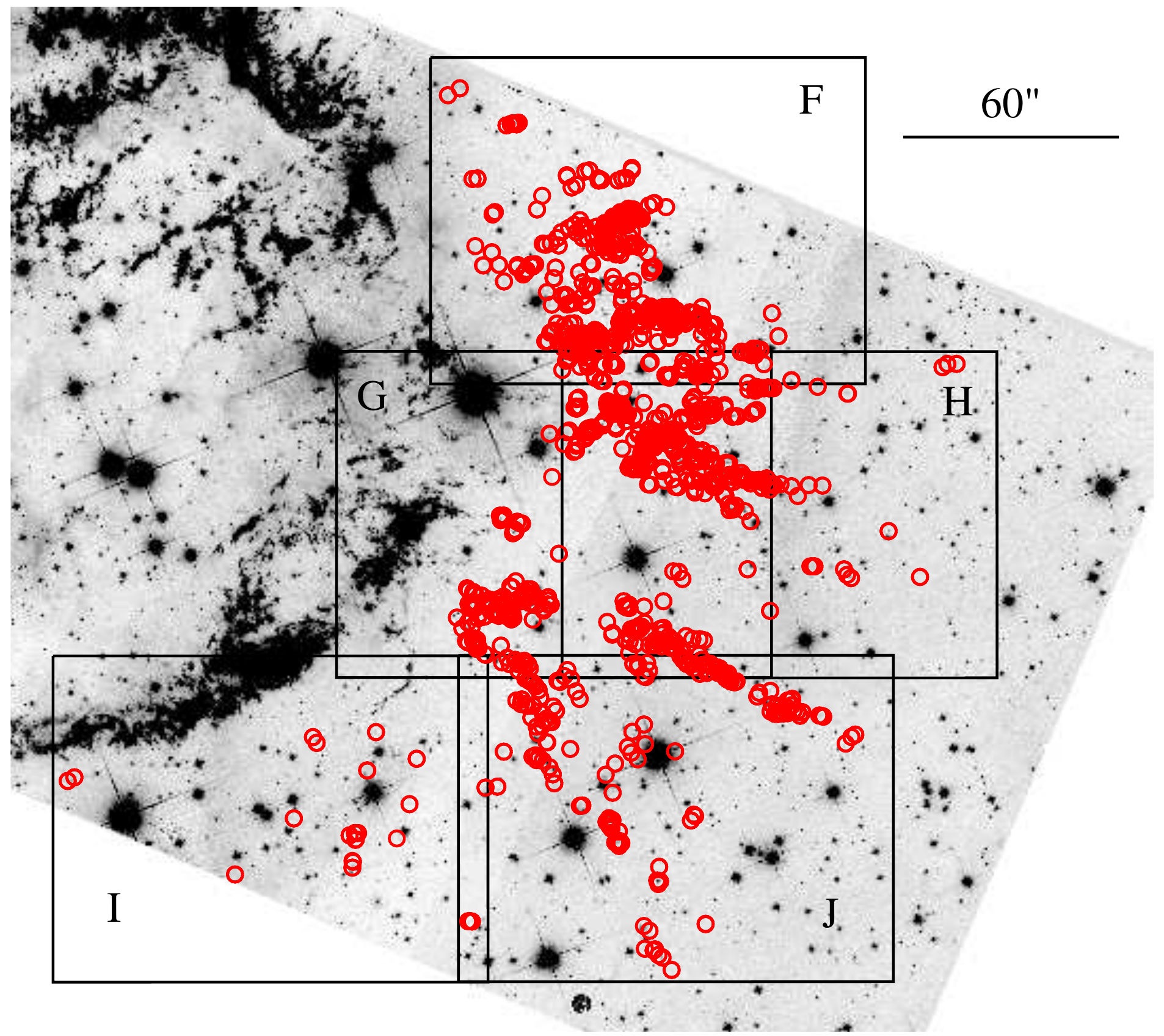}
        \caption{WFC3/IR F098M image of the eastern (left) and western (right) limbs
         of Cas A obtained in December 2011 with
        regions (A thru J) where enlargements are shown in Figures X - Y.
         }
        \label{fig:chart}
\end{figure*}

\newpage


\begin{figure*}[t]
        \centering
        \includegraphics[scale=.75]{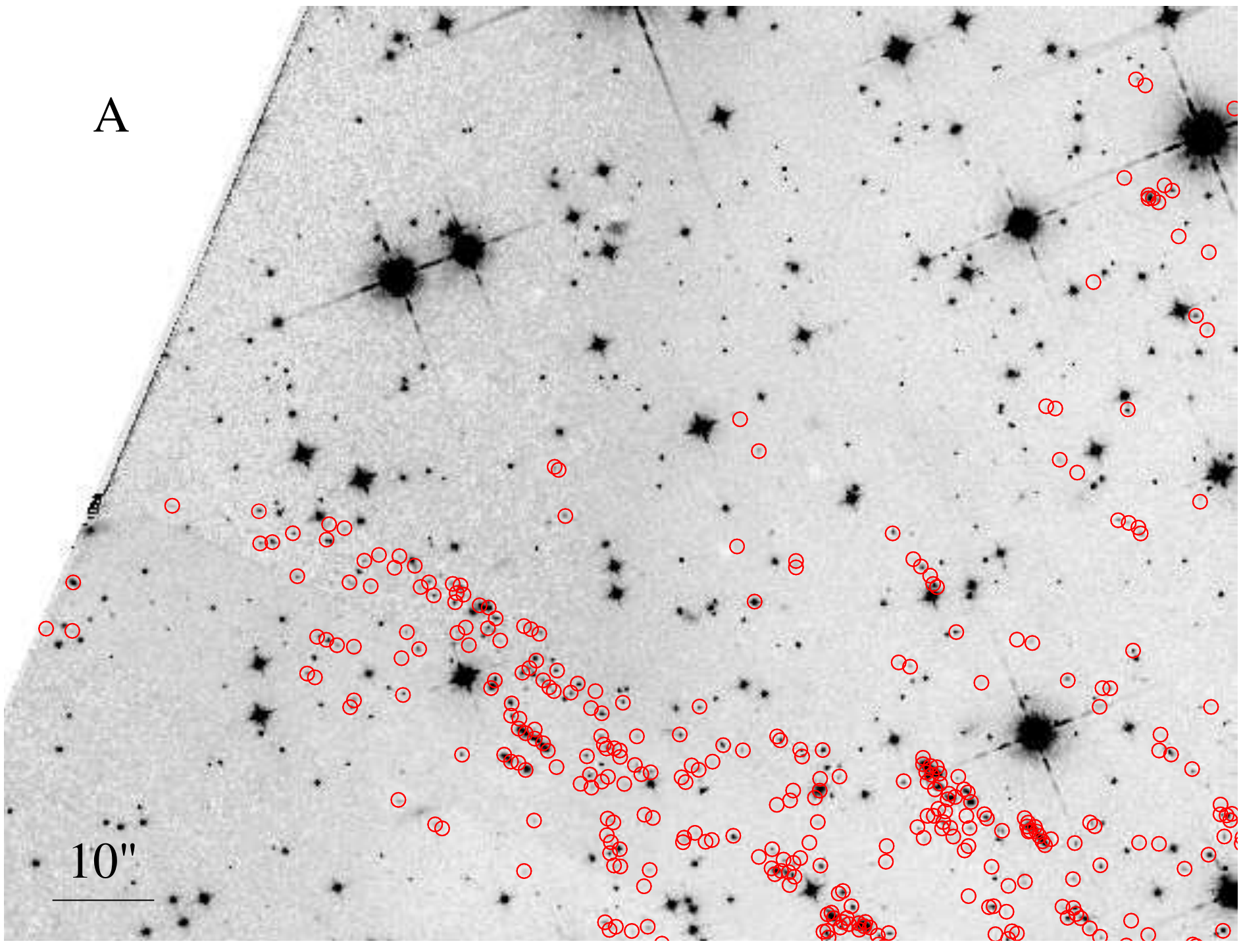}
        \includegraphics[scale=.75]{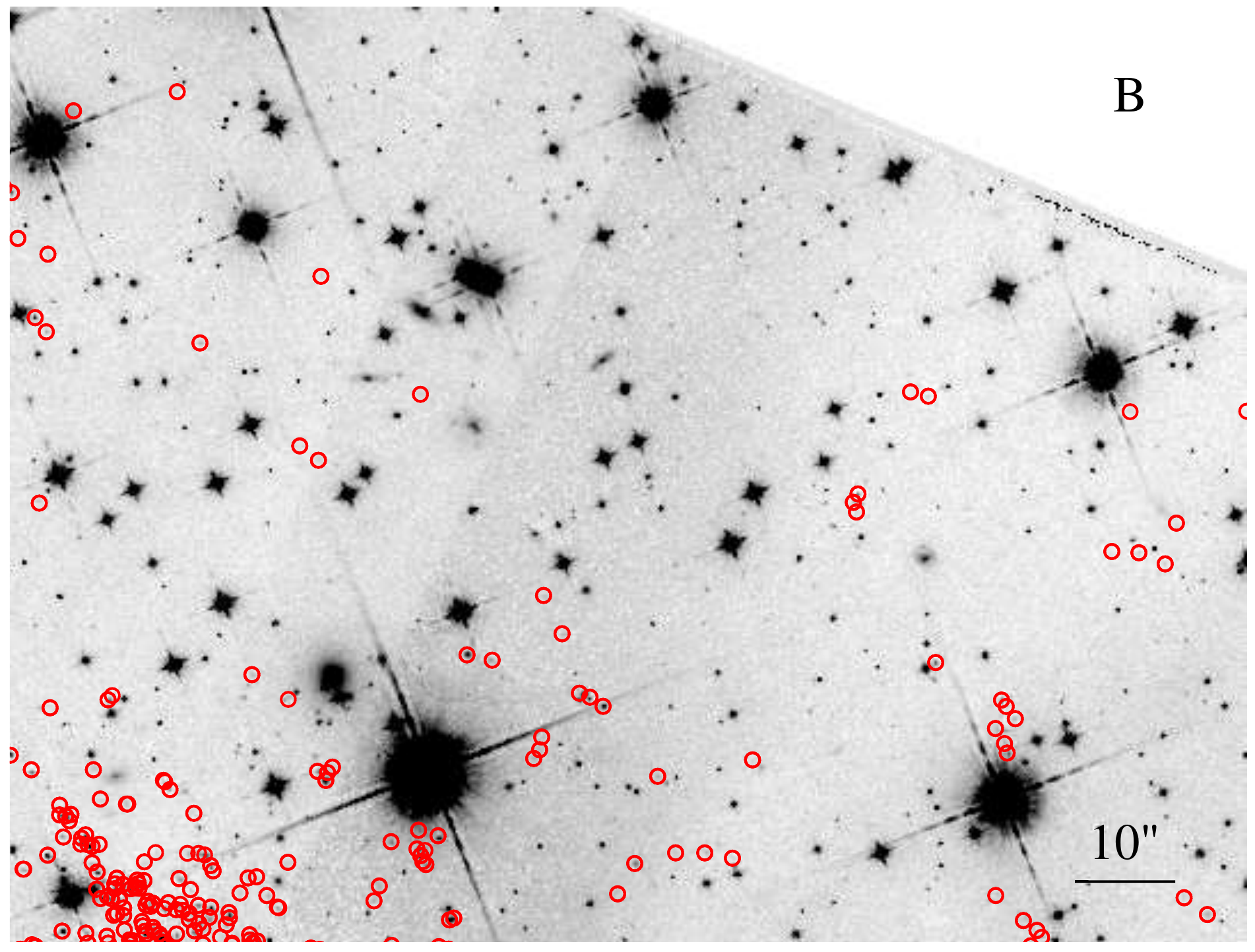}
\caption {Finding chart for ejecta knots in Regions A and B.  }
\label{regions_A_n_B}
\end{figure*}


\begin{figure*}[t]
        \centering
        \includegraphics[scale=.75]{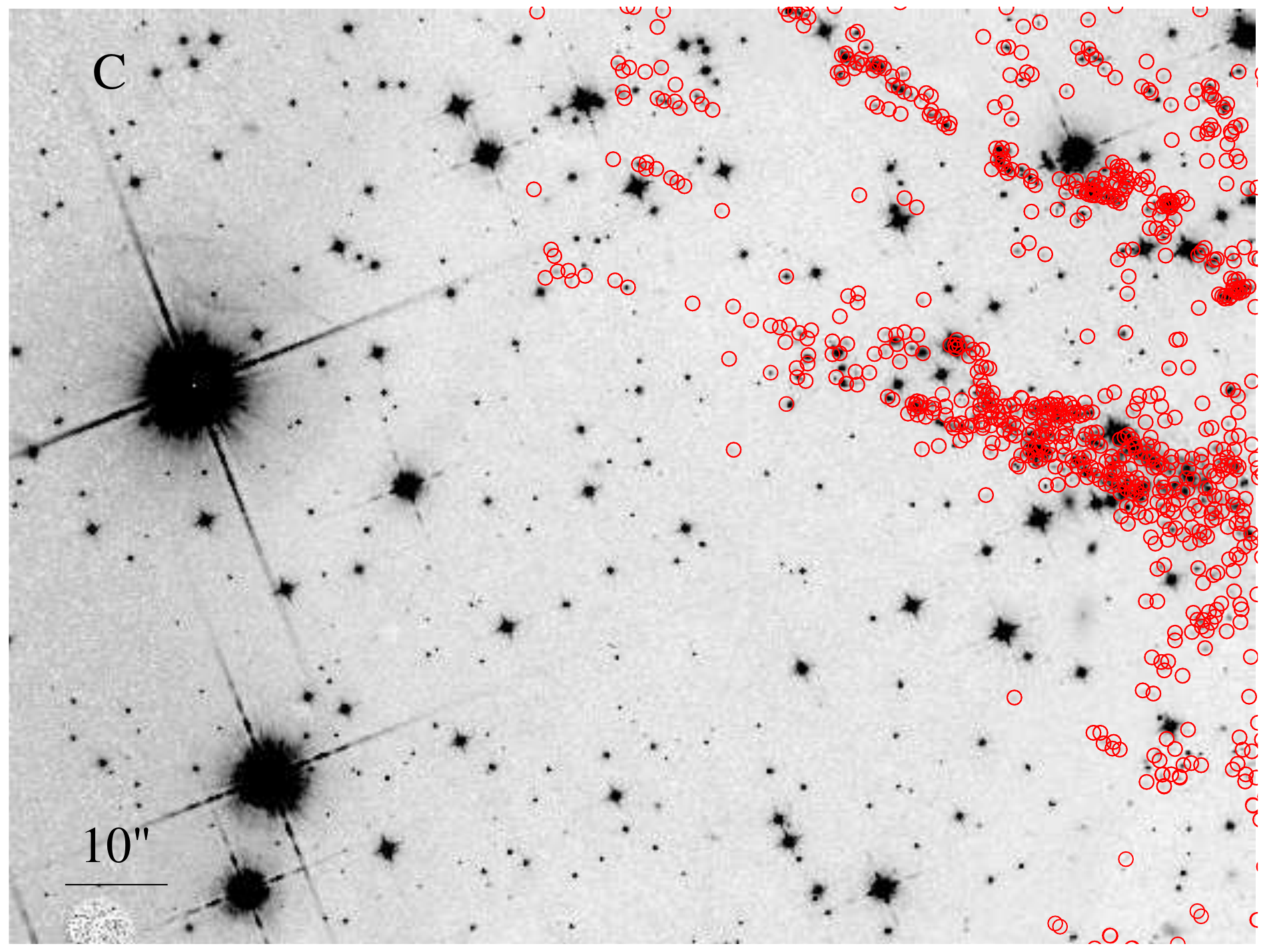}
        \includegraphics[scale=.75]{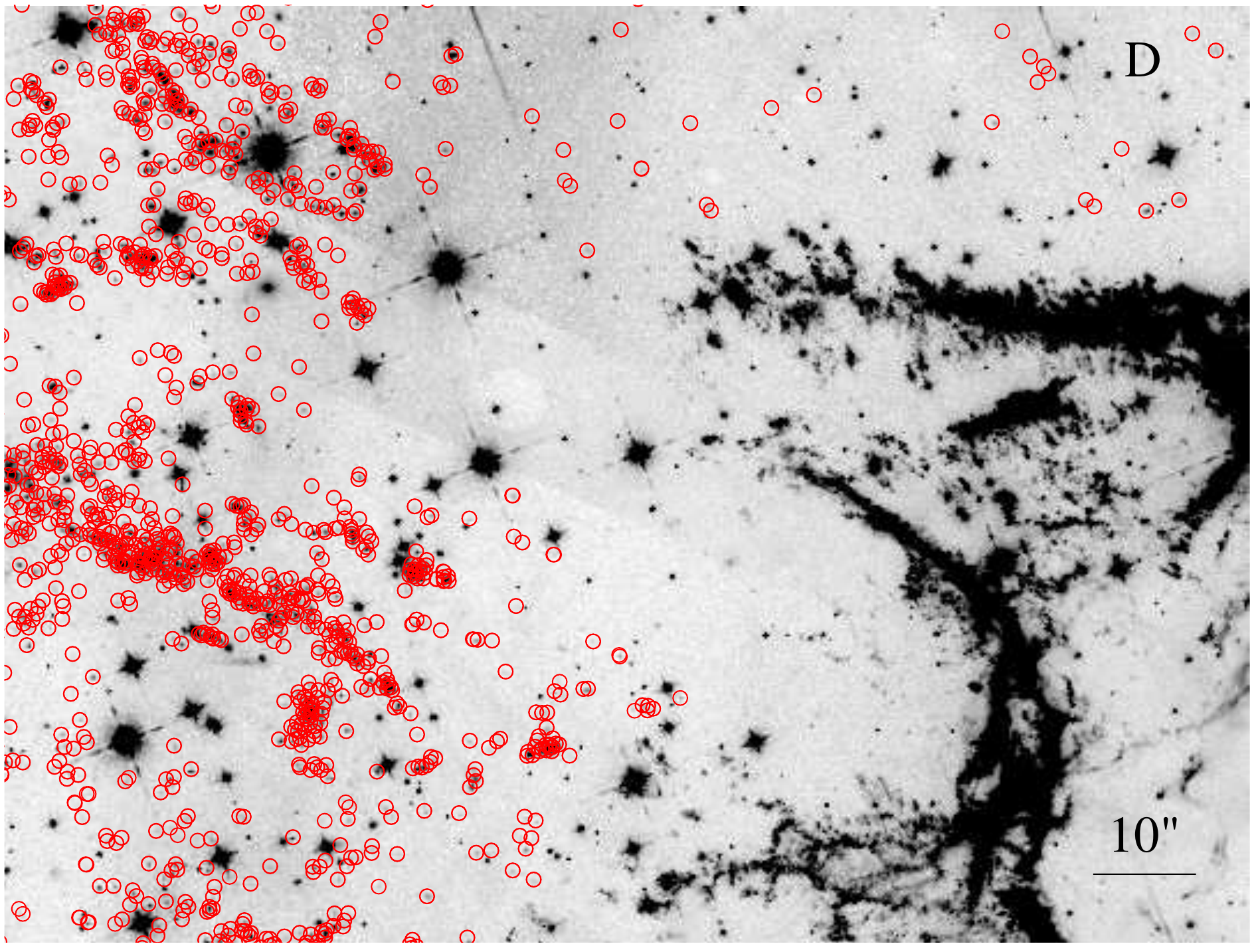}
\caption{ Finding chart for ejecta knots in Regions C and D. }
\label{regions_C_n_D}
\end{figure*}


\begin{figure*}[t]
        \centering
        \includegraphics[scale=.75]{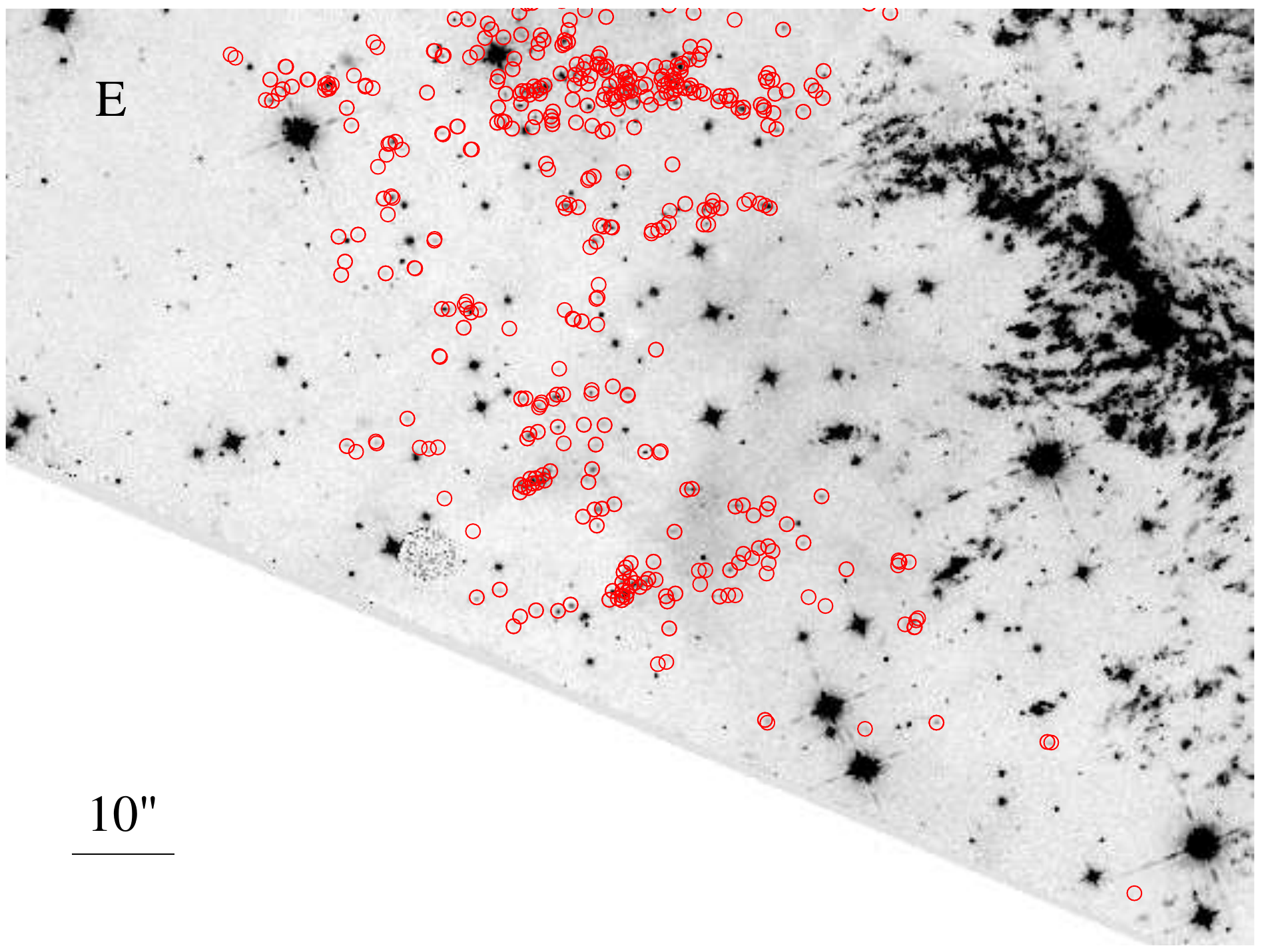}
        \includegraphics[scale=.75]{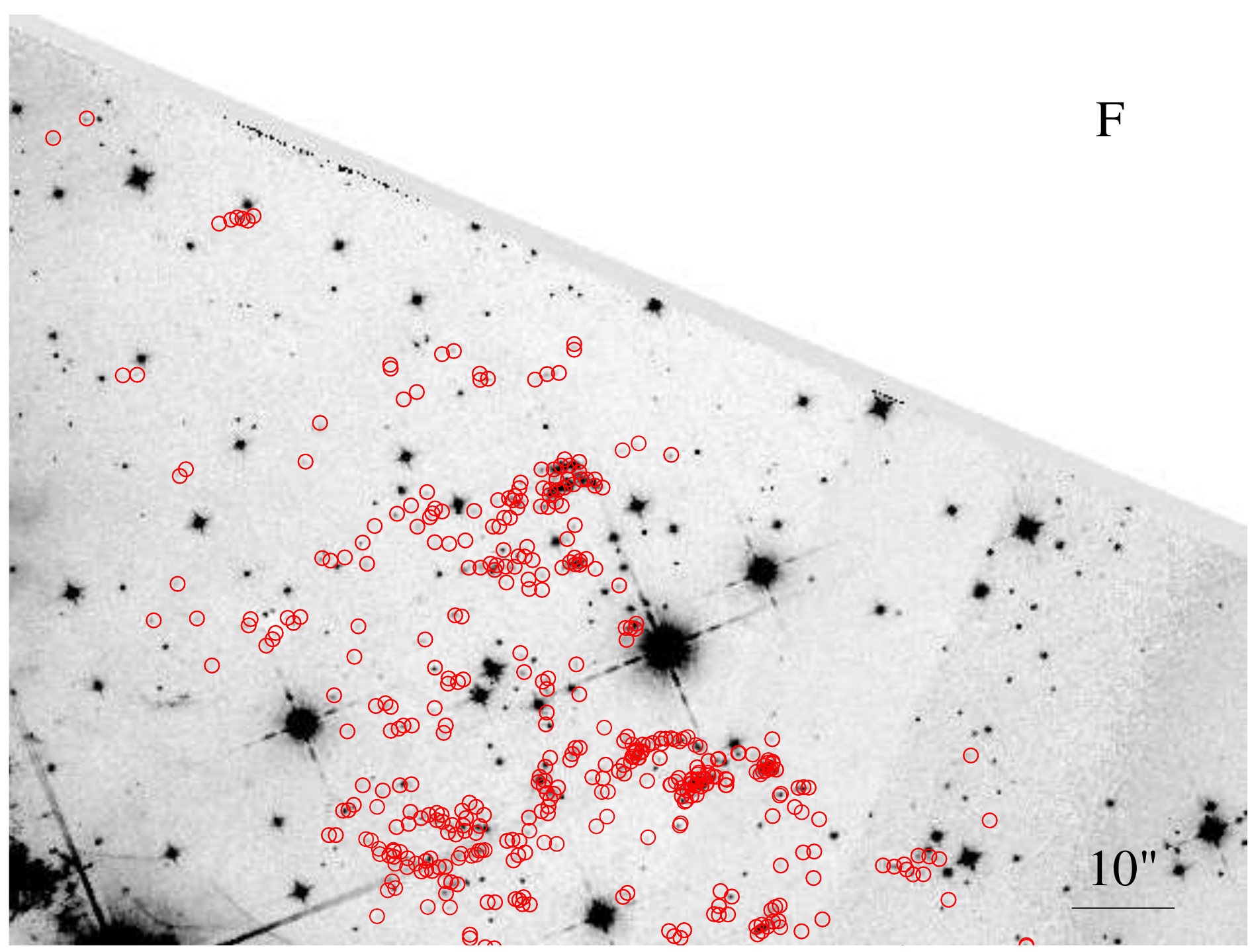}
\caption{Finding chart for ejecta knots in Regions E and F. } 
\label{regions_E_n_F}
\end{figure*}


\begin{figure*}[t]
        \centering
        \includegraphics[scale=.75]{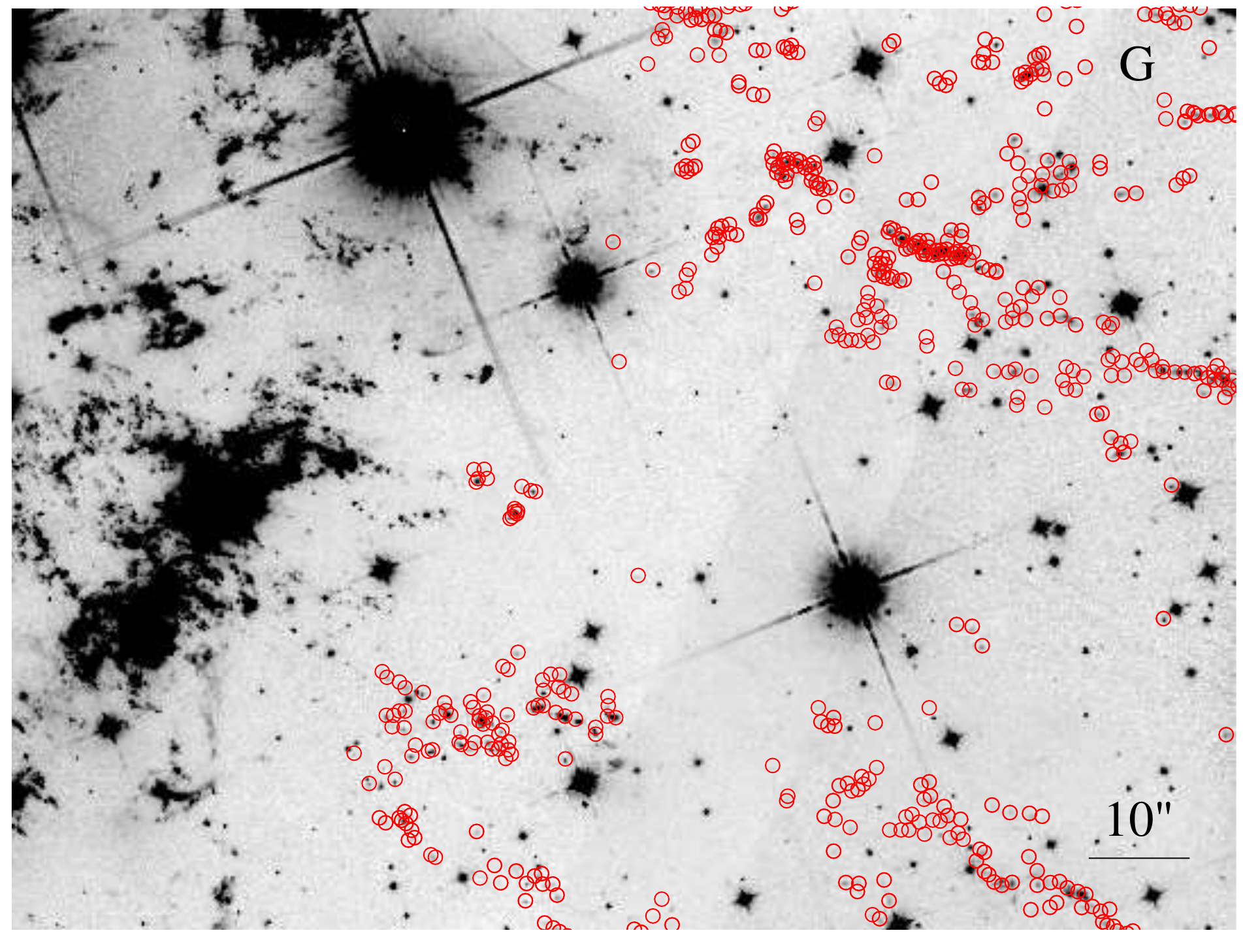}
        \includegraphics[scale=.75]{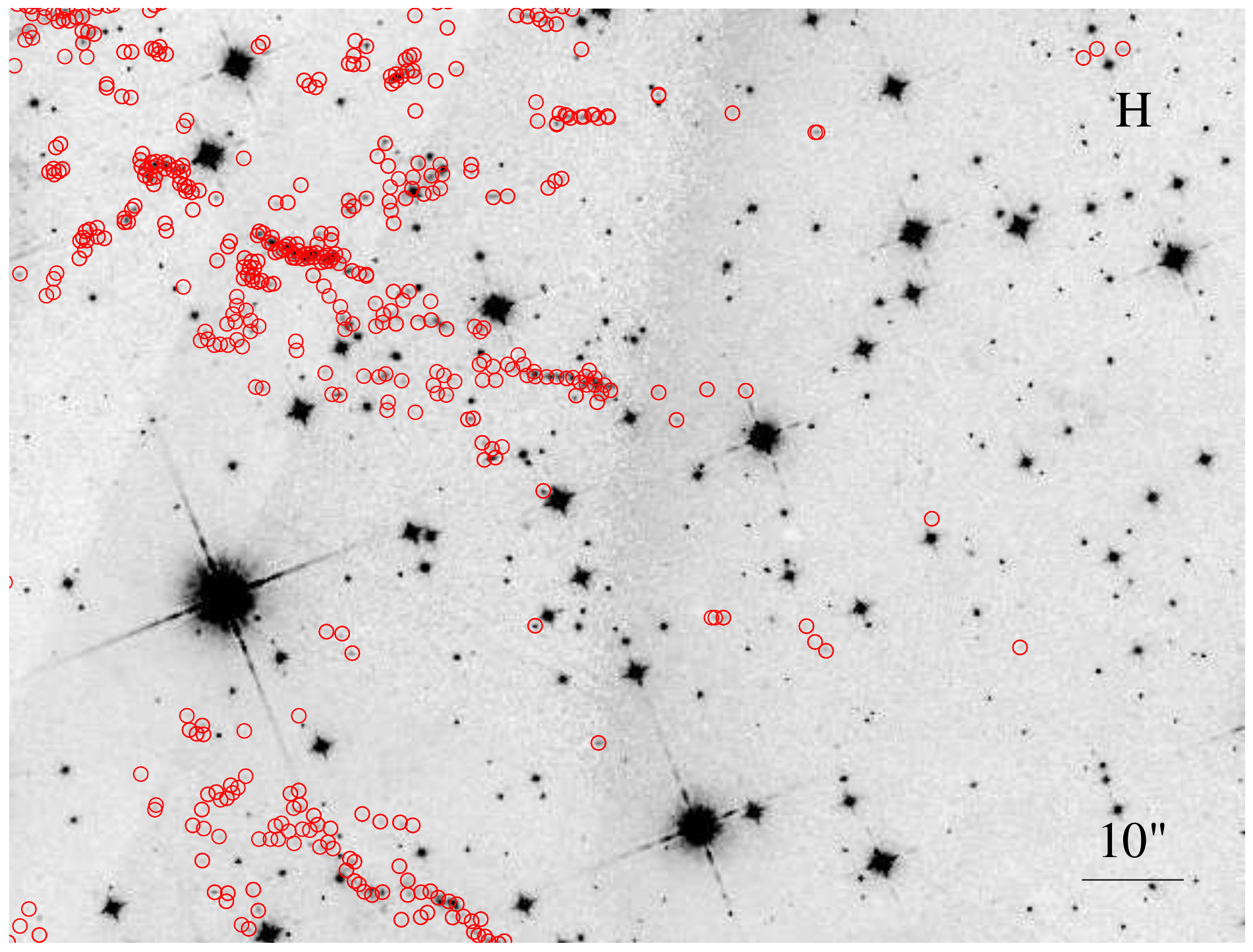}
\caption{Finding chart for ejecta knots in Regions G and H. }
\label{regions_G_n_H}
\end{figure*}


\begin{figure*}[t]
        \centering
        \includegraphics[scale=.75]{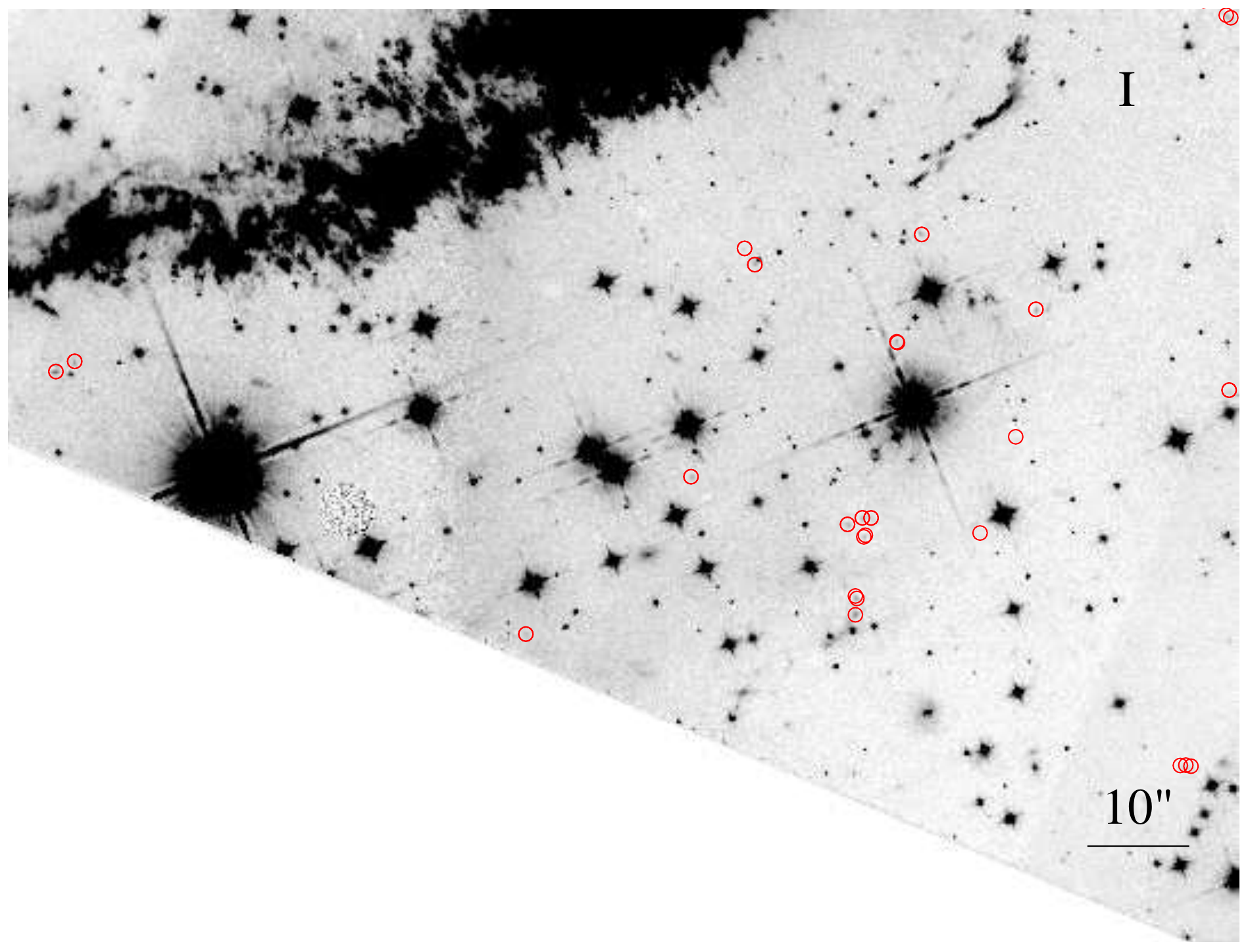}
        \includegraphics[scale=.75]{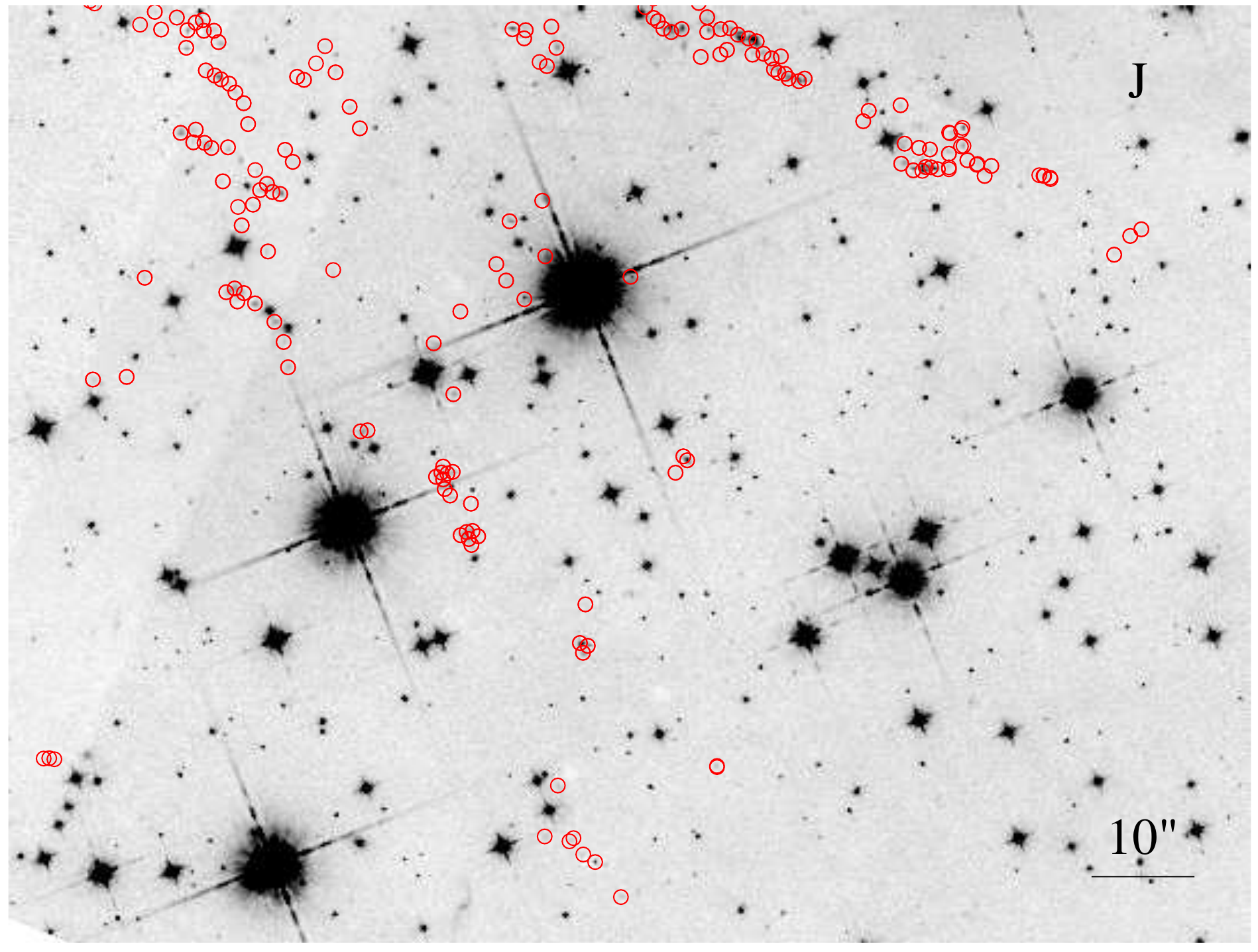}
\caption{Finding chart for ejecta knots in Regions I and J.  }
\label{regions_I_n_J}
\end{figure*}



\begin{figure*}[t]
        \centering
        \includegraphics[scale=.95]{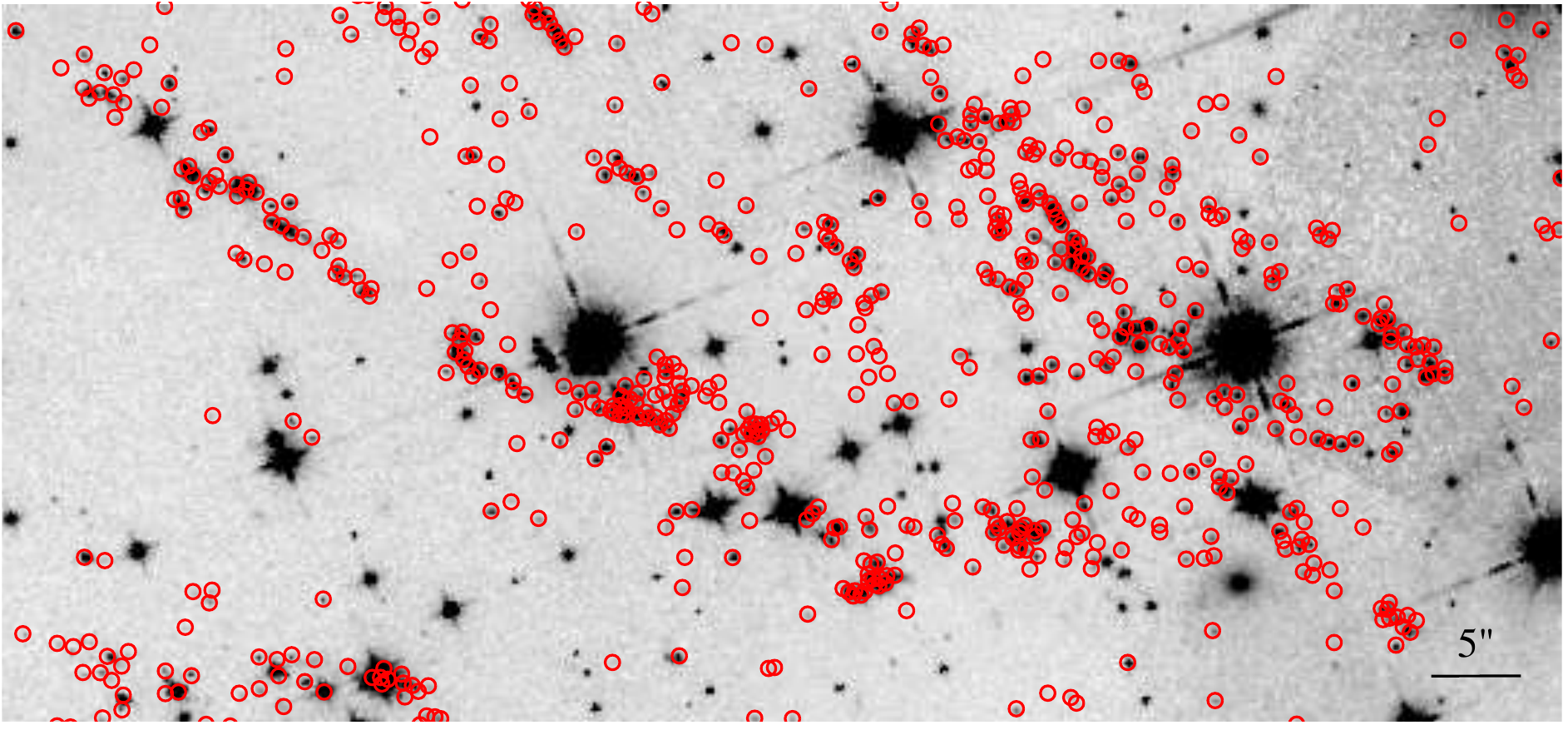}
        \includegraphics[scale=.95]{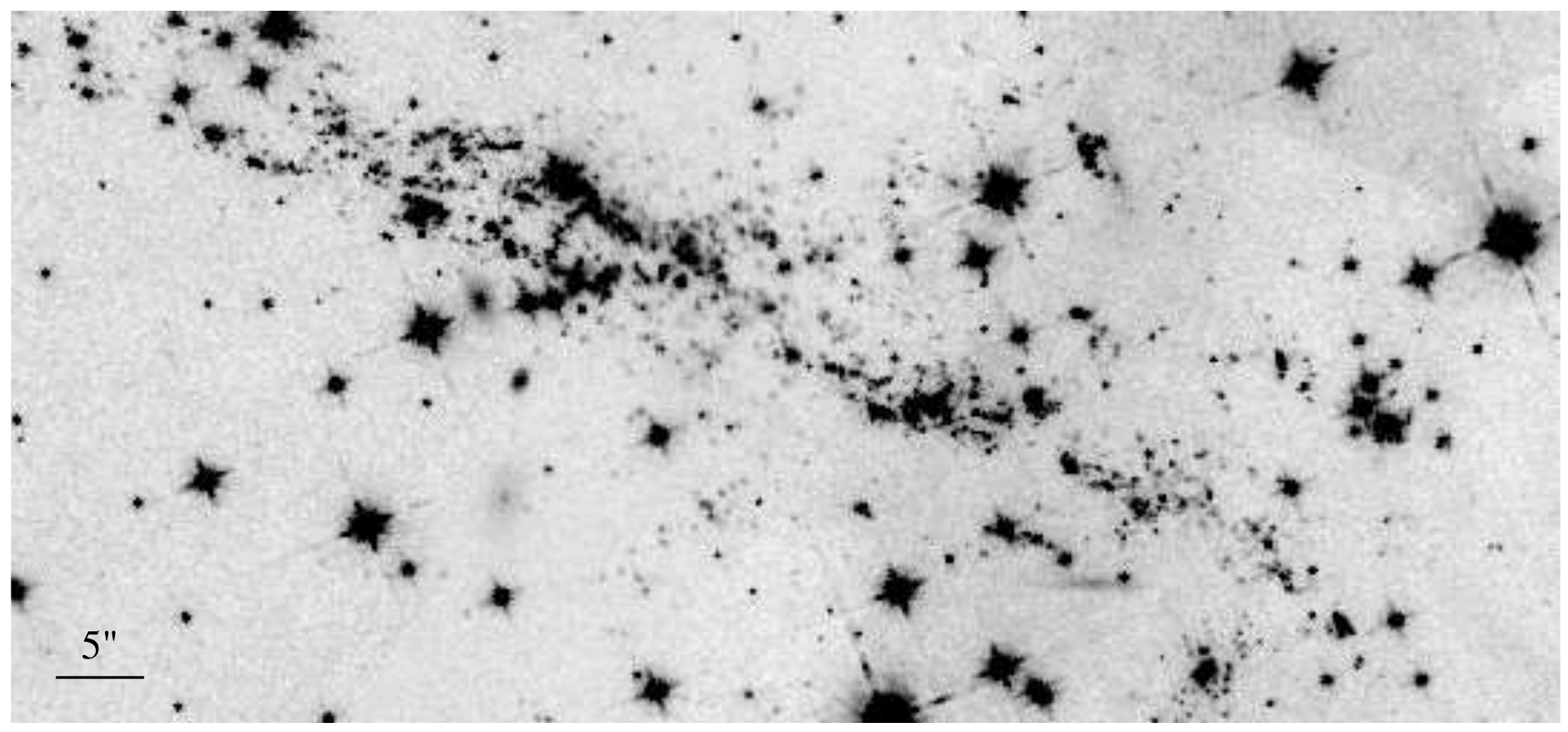}
\caption{Enlargement of the NE jet.  North up, East to the left. 
Approximate center of image is: $23^{\rm h} 23^{\rm m} 51.133^{\rm s}, +58\degr 50' 25\farcs3$ (J2000). 
          }
\label{NE_blowup1}
\end{figure*}

\begin{figure*}[t]
        \centering
        \includegraphics[scale=.95]{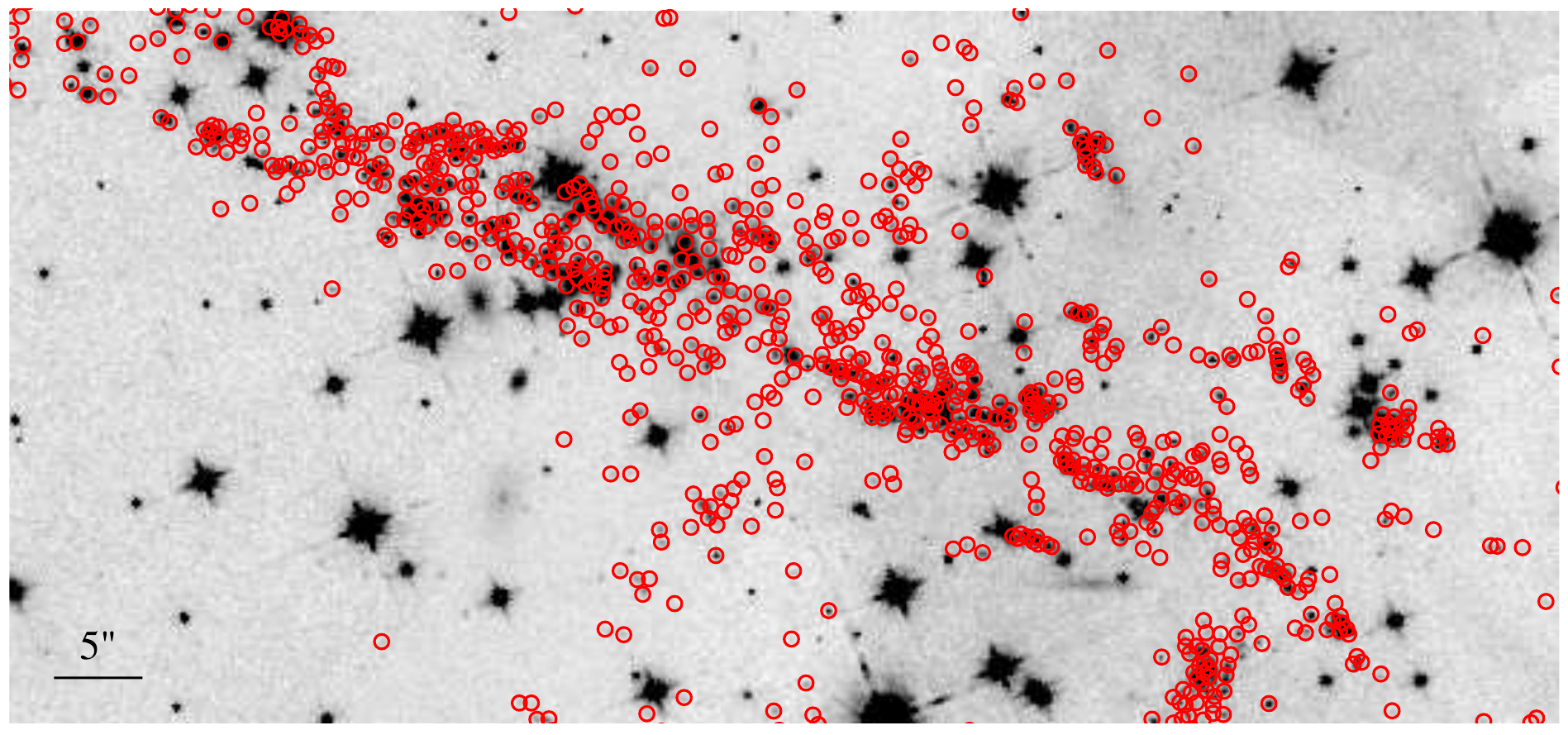}
        \includegraphics[scale=.95]{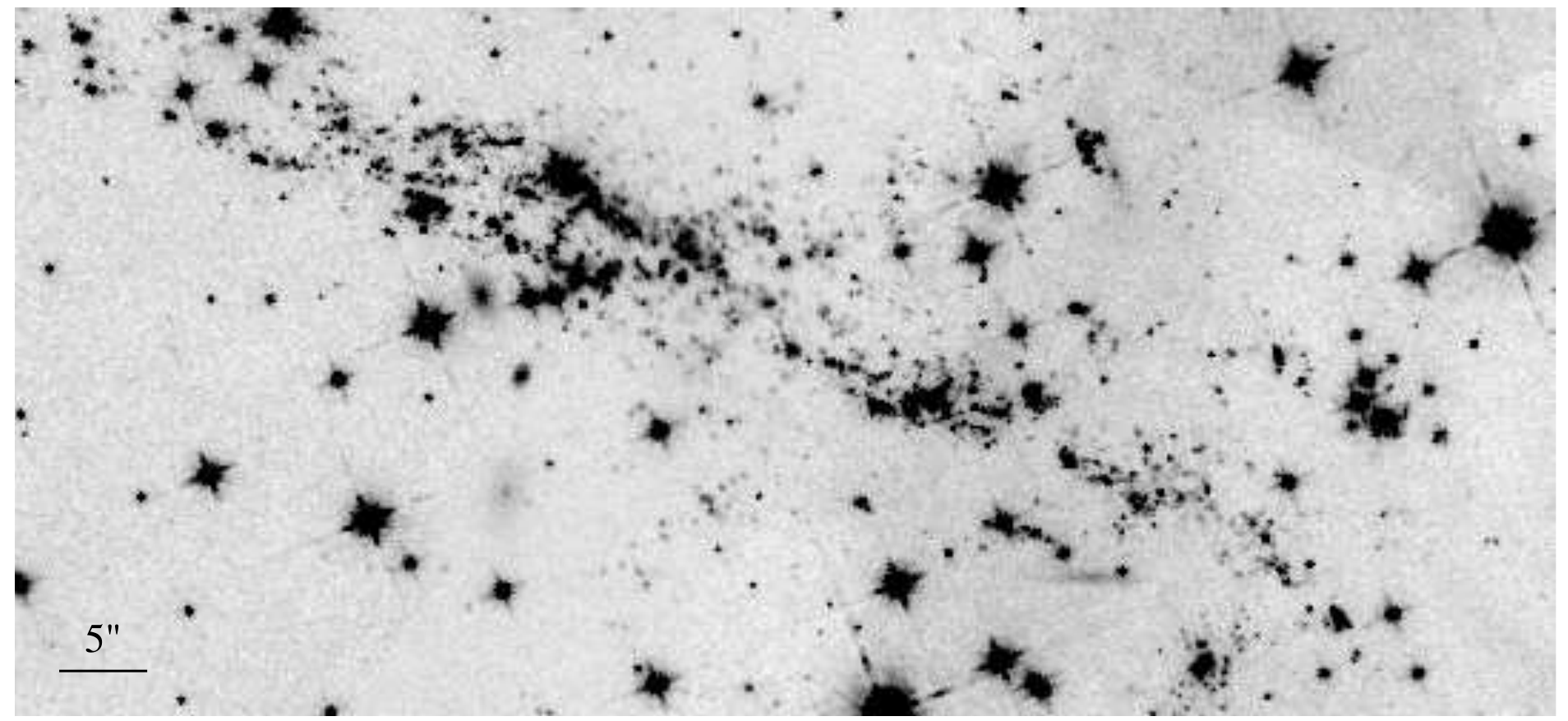}
\caption{Enlargement of the NE jet.  North up, East to the left.
Approximate center of image is: $23^{\rm h} 23^{\rm m} 50.352^{\rm s}, +58\degr 49' 48\farcs9$ (J2000). 
          }
\label{NE_blowup2}
\end{figure*}

\begin{figure*}[t]
        \centering
        \includegraphics[scale=.95]{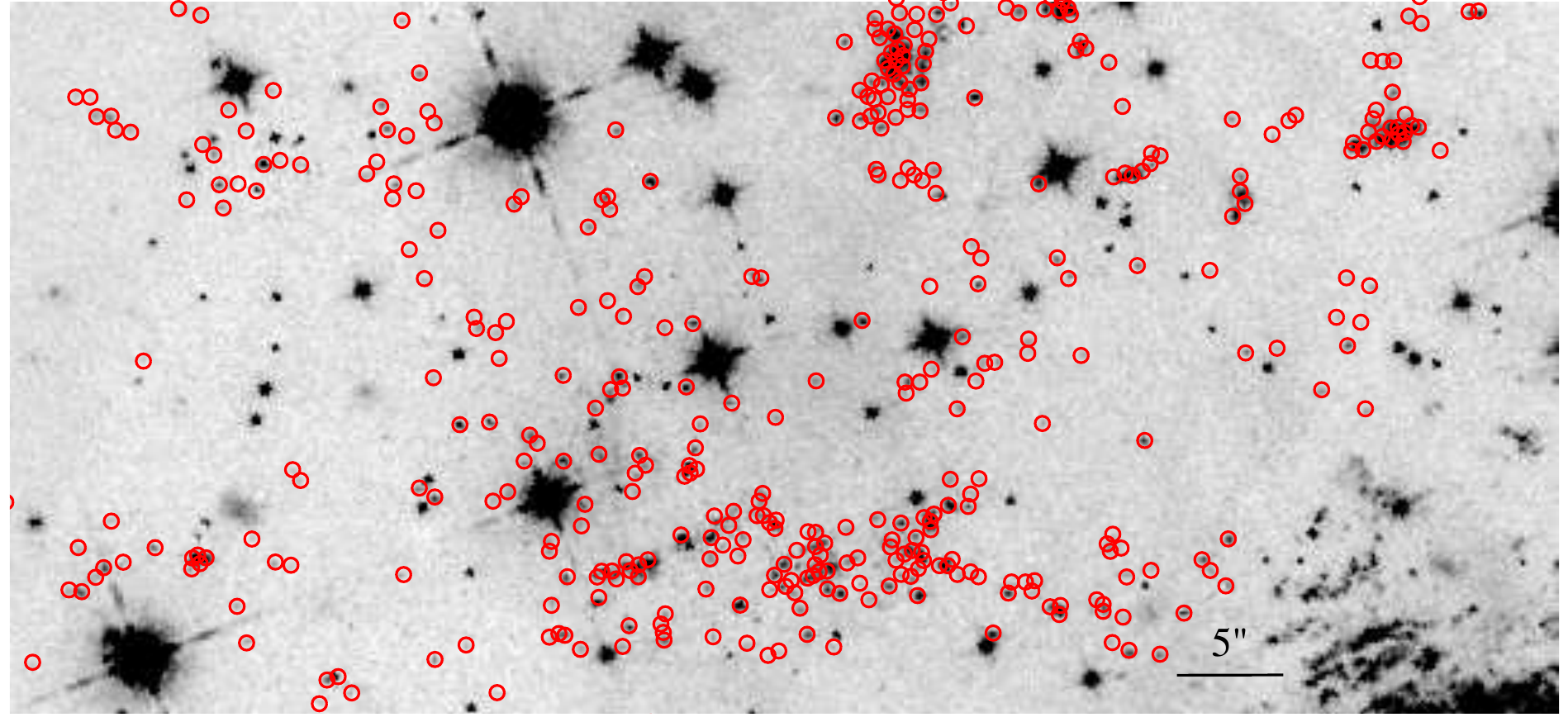}
        \includegraphics[scale=.95]{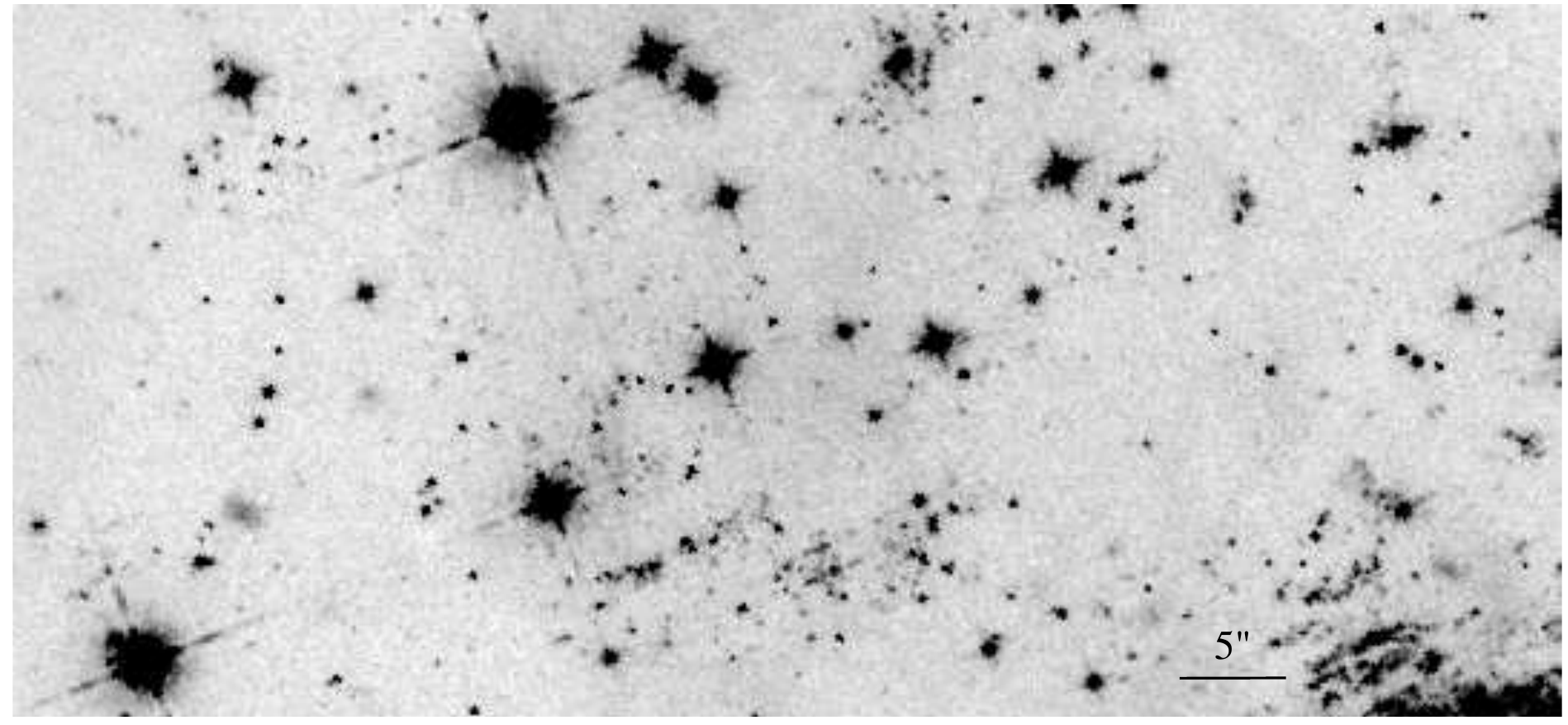}
\caption{ Enlargement of the NE jet.  North up, East to the left.
Approximate center of image is: $23^{\rm h} 23^{\rm m} 47.967^{\rm s}, +58\degr 49' 16\farcs2$ (J2000).
          }
\label{NE_blowup3}
\end{figure*}


\begin{figure*}[t]
        \centering
        \includegraphics[scale=.95]{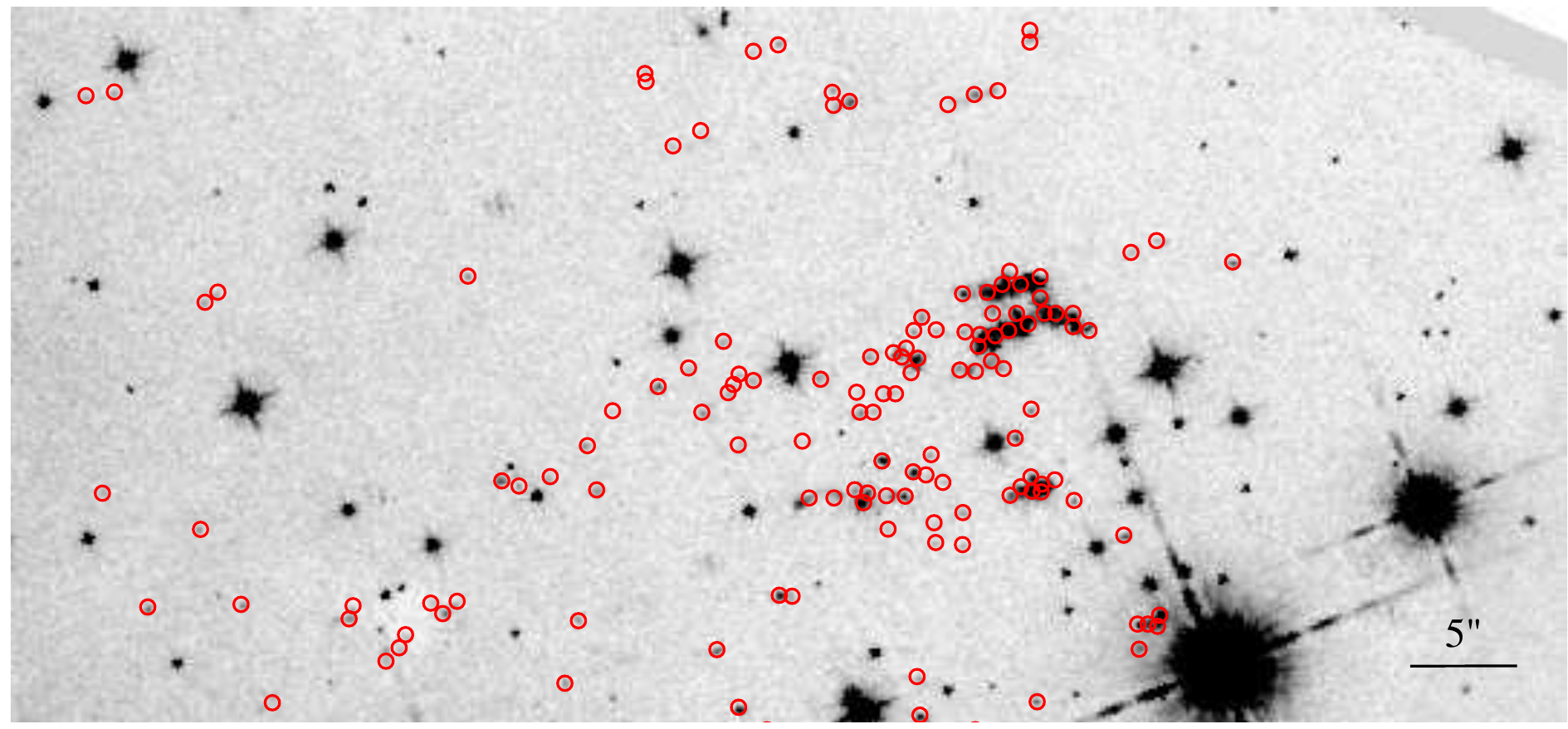}
        \includegraphics[scale=.95]{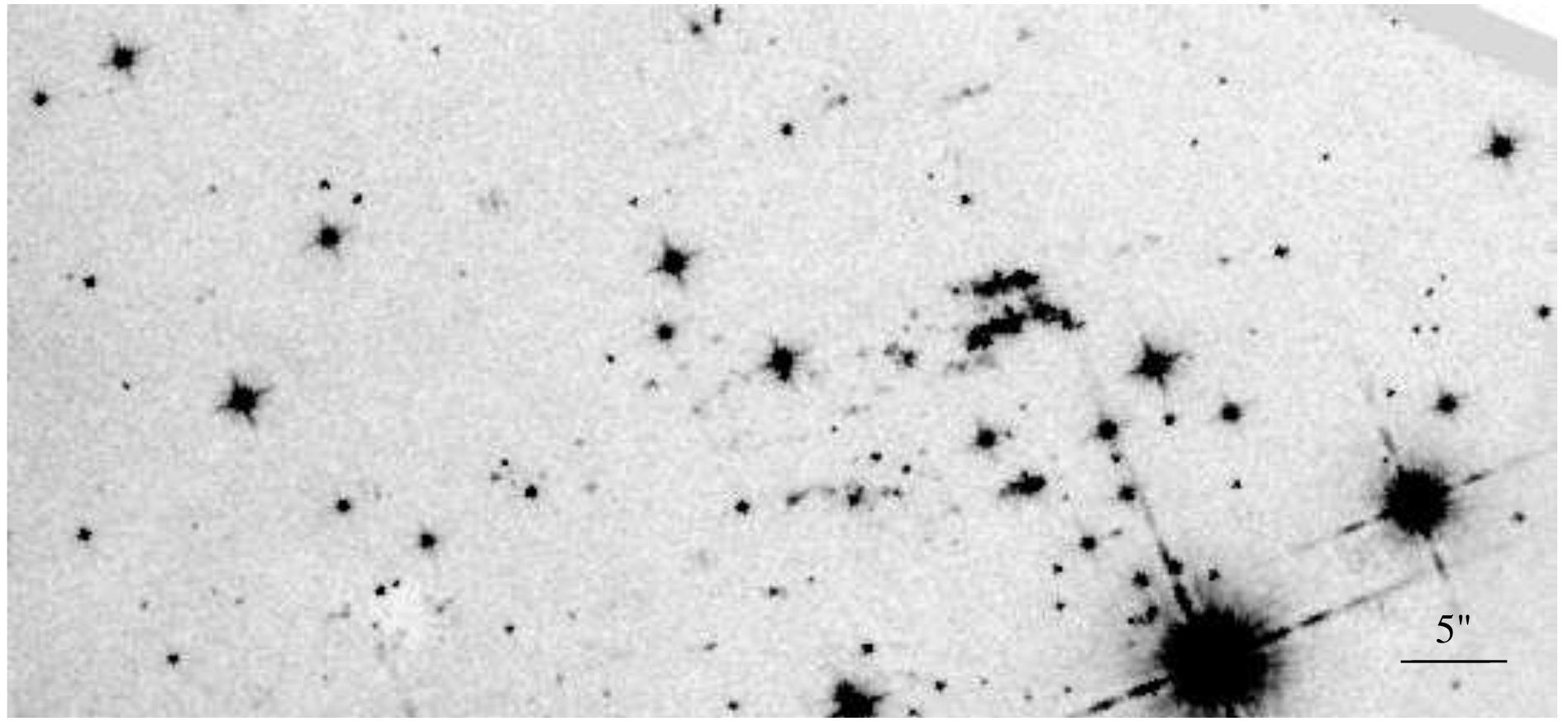}
\caption{Enlargement of the SW jet.  North up, East to the left.
Approximate center of image is: $23^{\rm h} 23^{\rm m} 6.430^{\rm s}, +58\degr 49' 32\farcs4$ (J2000).
}
\label{SW_blowup1}
\end{figure*}

\begin{figure*}[t]
        \centering
        \includegraphics[scale=.95]{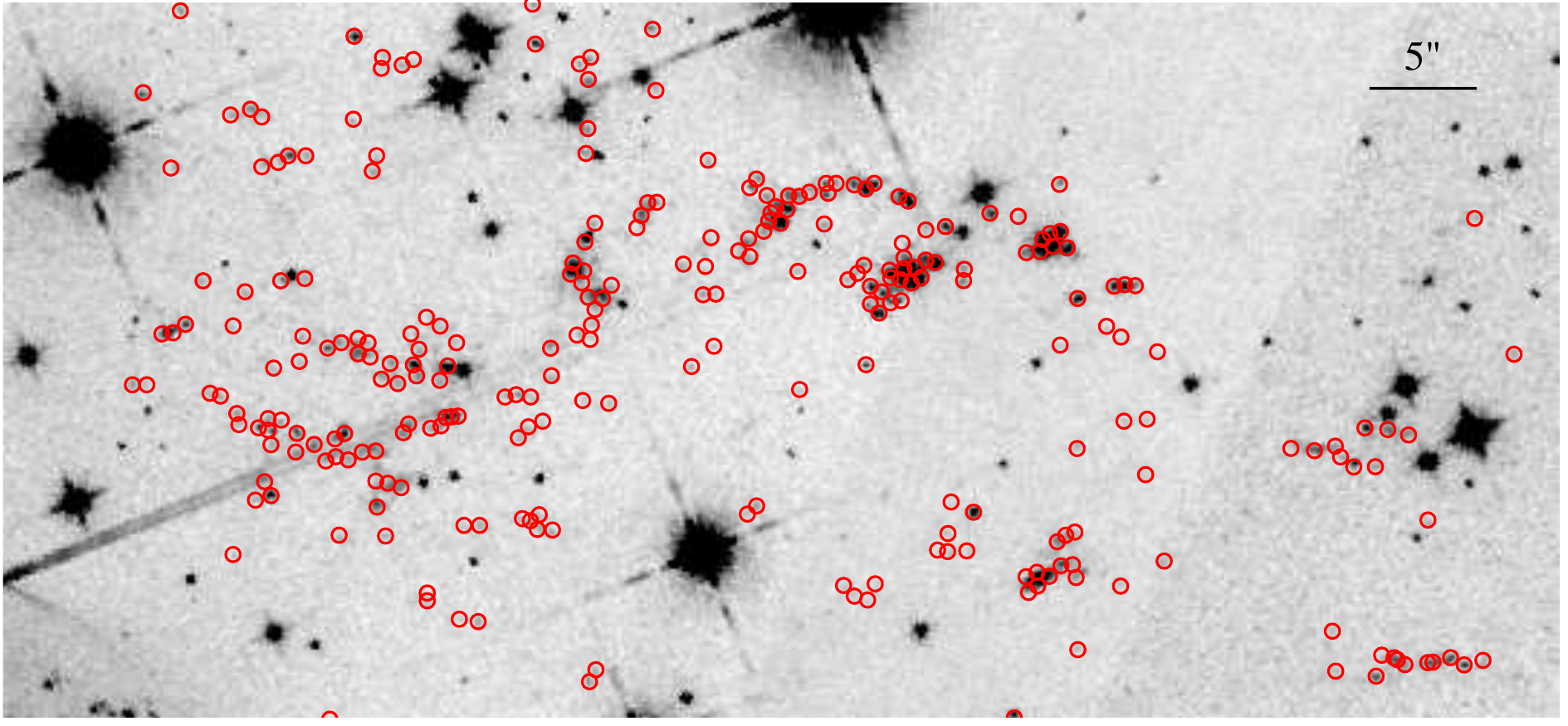}
        \includegraphics[scale=.95]{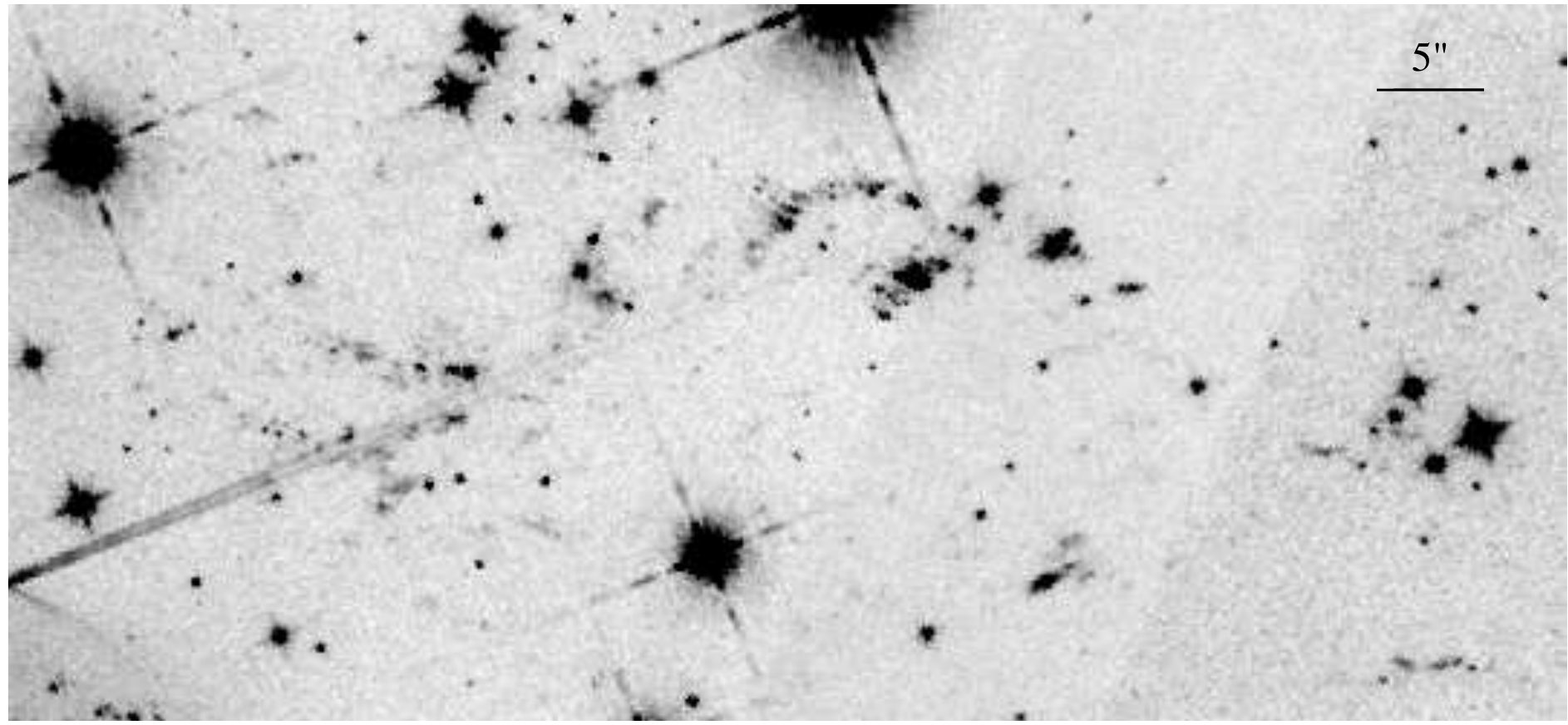}
\caption{ Enlargement of the SW jet.  North up, East to the left.
Approximate center of image is: $23^{\rm h} 23^{\rm m} 4.135^{\rm s}, +58\degr 49' 1\farcs2$ (J2000).
          }
\label{SW_blowup2}
\end{figure*}

\begin{figure*}[t]
        \centering
        \includegraphics[scale=.95]{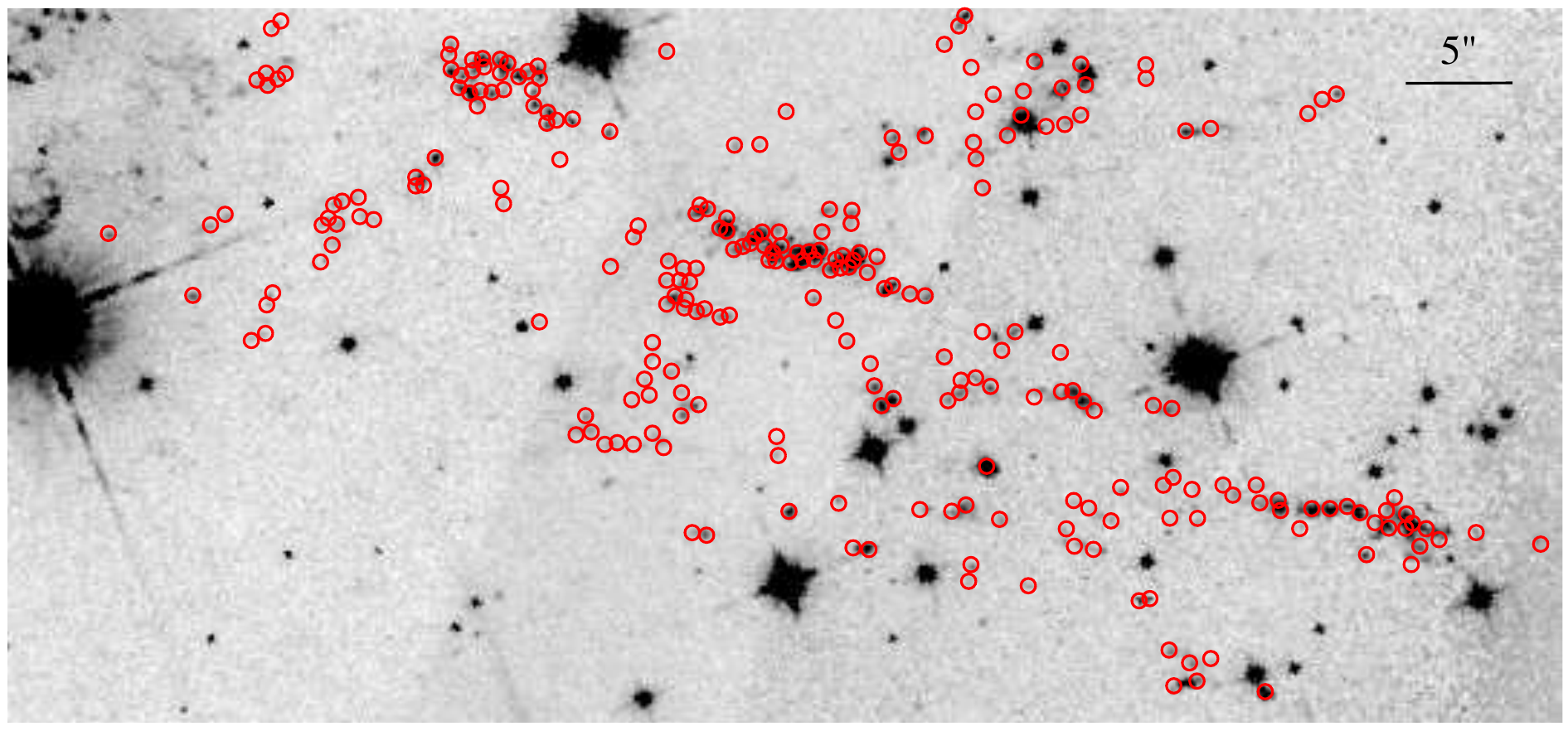}
        \includegraphics[scale=.95]{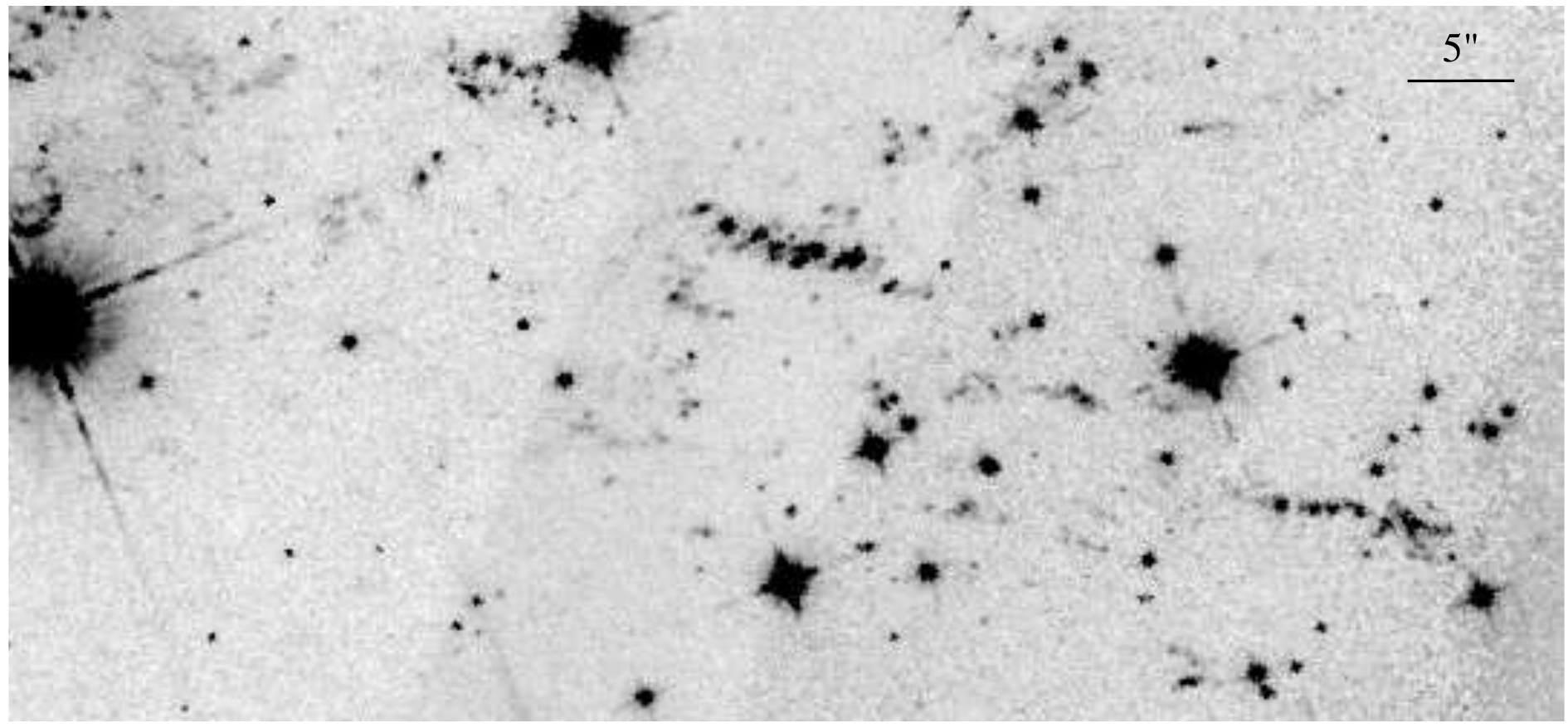}
\caption{Enlargement of the SW jet.  North up, East to the left.
Approximate center of image is: $23^{\rm h} 23^{\rm m} 3.80^{\rm s}, +58\degr 48' 28\farcs4$ (J2000).
         }

\label{SW_blowup3}
\end{figure*}

\begin{figure*}[t]
        \centering
        \includegraphics[scale=.95]{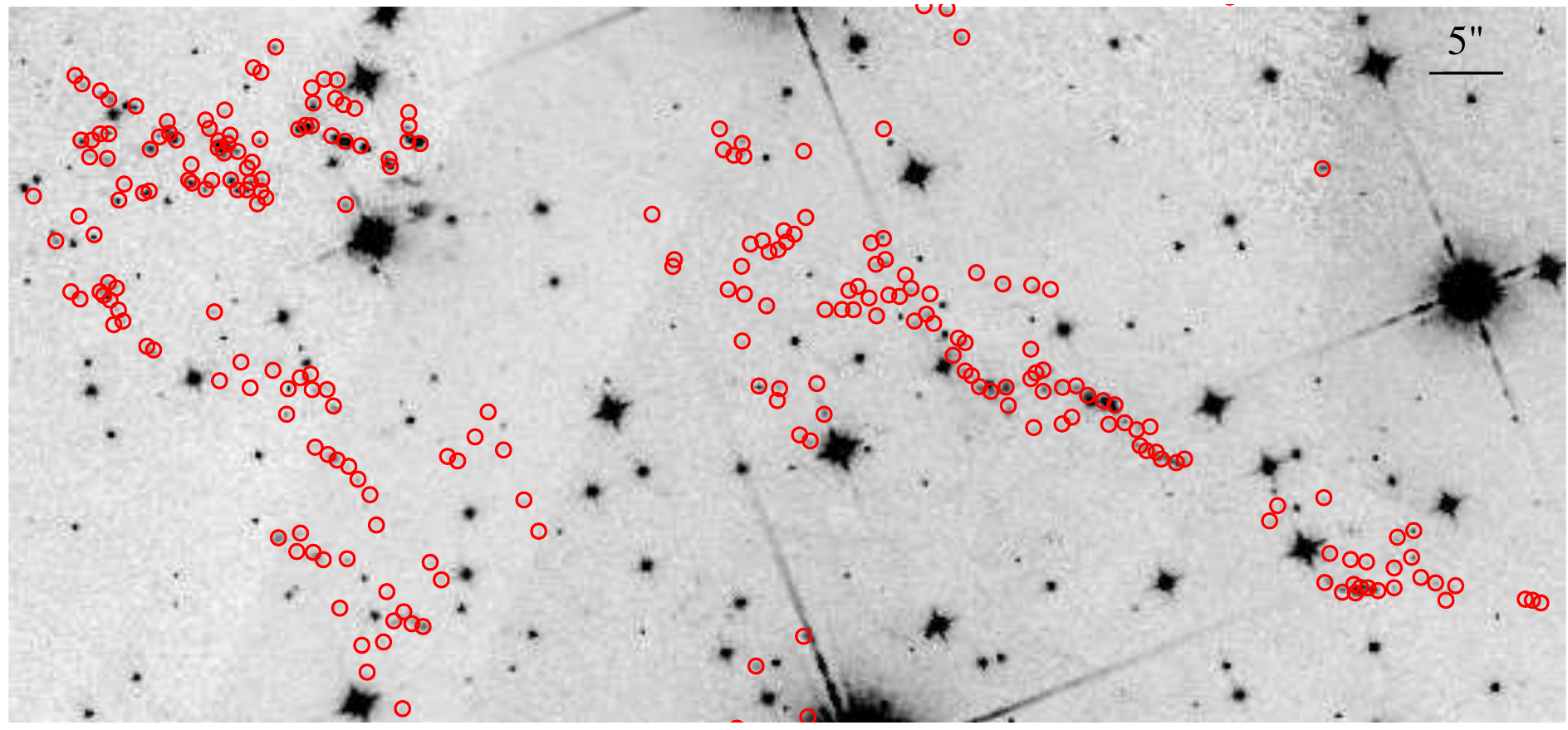}
        \includegraphics[scale=.95]{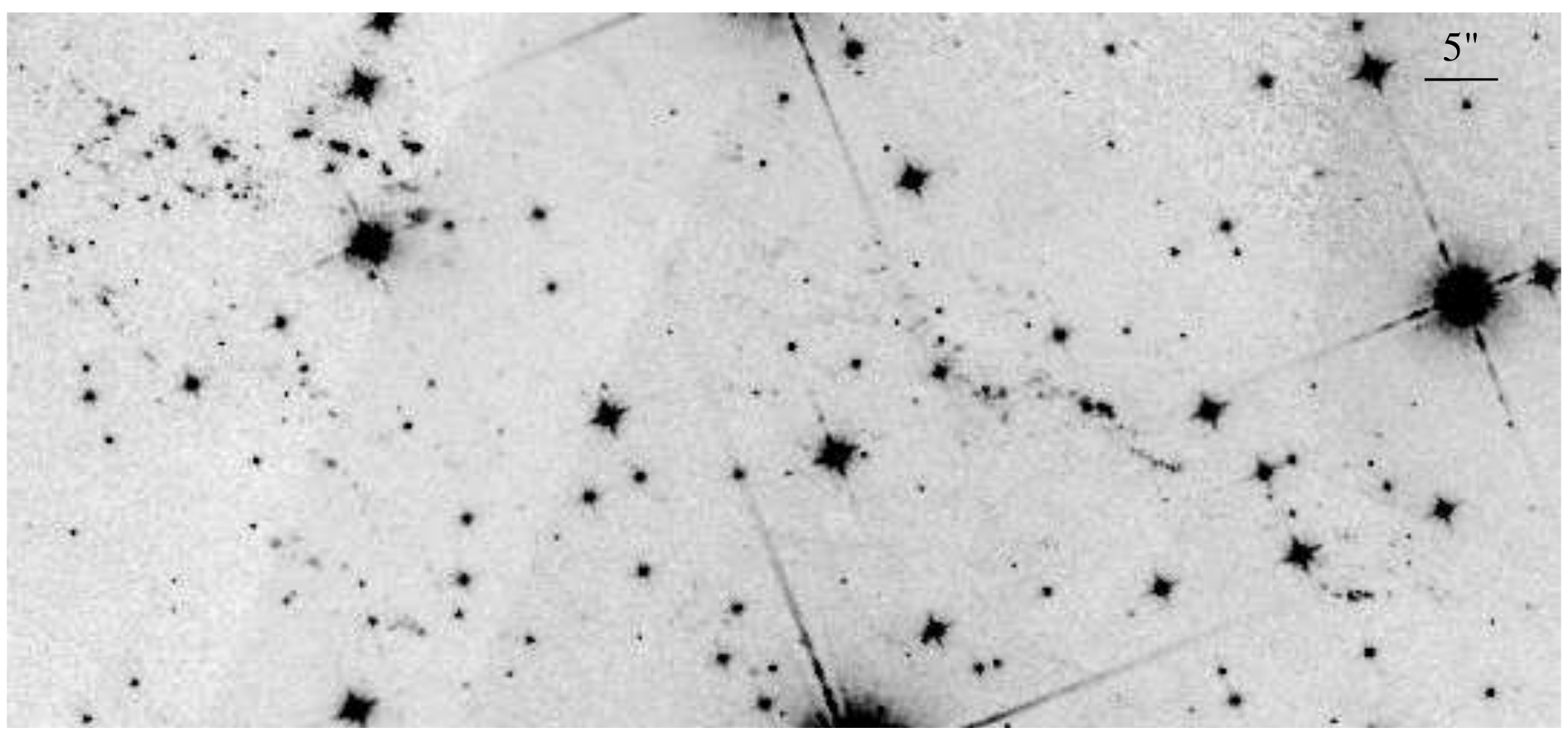}
\caption{ Enlargement of the SW Jet.  North up, East to the left.
Approximate center of image is: $23^{\rm h} 23^{\rm m} 4.70^{\rm s}, +58\degr 47' 32\farcs6$ (J2000).
          }
\label{SW_blowup4}
\end{figure*}

\end{document}